\documentclass[a4paper,11pt]{article}
\usepackage{pos}

\usepackage{amsmath}
\usepackage{amsfonts}
\usepackage{amsthm}
\usepackage{fixmath}
\usepackage{braket}
\usepackage{slashed}
\usepackage{fontenc}
\usepackage{bm}
\usepackage{xfrac}
\usepackage{mathtools}
\usepackage{lineno}
\renewcommand{\bm}{\mathbold}
\newcommand{\mr}{\mathrm}

\newcommand{\mc}{\mathcal}
\def\ri{\mr{i}} 
\def\tr{\mr{tr}~}

\newcommand{\as}{\alpha_s}
\newcommand{\md}{m_D}

\title{Heavy quarks at finite temperature}
\author*[a]{Johannes Heinrich Weber}

\affiliation[a]{Institut f\"ur Physik \& IRIS Adlershof \& RTG 2575, 
Humboldt-Universit\"at zu Berlin, %\\
  D-12489 Berlin, Germany}

\emailAdd{johannes.weber@physik.hu-berlin.de}

\abstract{

New incarnations of heavy-ion collision experiments are turning our 
attention to hard processes and a more fine-grained resolution of the QGP. 
In this endeavor quarkonia or open heavy flavors turn out to be versatile 
probes, which are usually described through models based on perturbative 
QCD, AdS, and effective field theories. The lattice provides 
nonperturbative input and constraints to such models.

In-medium bottomonia, the complex static quark-antiquark potential, and 
the heavy-quark momentum diffusion coefficient are key quantities where 
lattice gauge theory has recently achieved significant progress with impact 
for heavy-ion phenomenology.
We review these lattice results, relate them to phenomenological 
applications, and close with an outlook.

}

\FullConference{%
 The 38th International Symposium on Lattice Field Theory, LATTICE2021
  26th-30th July, 2021
  Zoom/Gather@Massachusetts Institute of Technology
}

\begin{document}
\maketitle

\section{Introduction}

Both the asymptotic freedom at short distances, where the weak-coupling 
approach applies, as well as a wide range of strictly non-perturbative, 
mutually interrelated long-distance phenomena are well-established in 
experimental and theoretical studies of nuclear matter (see, e.g., 
Refs.~\cite{Akiba:2015jwa, Ding:2016qdj, Bazavov:2019lgz, Bazavov:2020teh} 
for a review). 
On the one hand, the asymptotic freedom predicts that nuclear matter can 
be understood at very high temperatures or densities as a hot plasma of 
its fundamental partons, which are interacting weakly~\cite{Gross:1980br}. 
This high-temperature state of nuclear matter is called the quark-gluon 
plasma (QGP), whose properties are determined by the interplay of its 
scales -- the temperature $T$, the Debye mass $m_D \sim gT$, and the 
magnetic scale $g^2T$ -- which are hierarchically ordered as 
$g^2T \ll m_D \ll \pi T$ in the strict weak-coupling limit.
On the other hand, the infrared regime of nuclear matter at sufficiently 
low temperatures and densities can be described to a good approximation 
as a gas of essentially non-interacting hadrons and resonances exhibiting 
their respective non-perturbative vacuum properties~\cite{Karsch:2003vd}. 
This low-temperature state of nuclear matter is called a hadron gas. 
As a description in terms of a hadron gas would lead to exponential growth 
in the density of states, it must eventually break down at high 
temperatures~\cite{Hagedorn:1965st}.

%%%%%%%%%%%%%%%%%%%%%%%%%%%%%%%%%%%%%%%%%%%%%%%%%%%%%%%%%%%

Instead, the properties of nuclear matter and the nature of its degrees of 
freedom undergo dramatic transitions when its temperature is increased 
towards and beyond certain thresholds that depend strongly on the details 
of the particle content, in particular on the number and mass of the light 
quark flavors~\cite{Cabibbo:1975ig}.  
On the one hand, the transition in the physical world, dominated by the 
nearby chiral transition at $T_\chi \approx 132\,\mathrm{MeV}$ for $N_f=2$ 
massless flavors and a physical strange quark~\cite{HotQCD:2019xnw}, is a 
smooth and broad crossover centered at the pseudocritical temperature of 
$T_\mathrm{pc}\approx 157\,\mathrm{MeV}$ with an uncertainty of about 
1\%~\cite{HotQCD:2018pds, Borsanyi:2020fev}. 
On the other hand, in the $\mathrm{SU}(3)$ pure gauge theory the 
deconfinement transition at $T_d \approx 270\,\mathrm{MeV}$ is the relevant 
threshold. 
Not too far above those respective thresholds, i.e. up to 
$T \sim 300\,\mr{MeV}$, the corresponding thermal medium produced in 
heavy-ion collisions behaves as an almost inviscid, nearly perfect fluid, 
which indicates a strongly-coupled medium, while quite a few of its 
properties become quantitatively similar to predictions from the 
weak-coupling expansion at $T > 300\,\mr{MeV}$, e.g., 
heavy quark-antiquark free energies in the static limit~\cite{Bazavov:2018wmo, Bazavov:2020teh}. 
At present, how and when this thermal medium becomes weakly coupled are 
unsolved questions, which are addressed in ongoing research.%~[\ldots]. 

%%%%%%%%%%%%%%%%%%%%%%%%%%%%%%%%%%%%%%%%%%%%%%%%%%%%%%%%%%%

Heavy quarks, either considered as isolated partons or within a heavy 
quark-antiquark pair, are almost ideal probes for the properties of the 
strong interactions 
(see, e.g., Refs.~\cite{Brambilla:2010cs, Rothkopf:2019ipj} for recent reviews). 
The former can be treated in many realistic scenarios as slowly-moving, 
non-relativistic probes with $m_h v \ll m_h$ in the fundamental representation 
of the gauge group, for whom particle-antiparticle creation and annihilation 
is strongly suppressed. 
When immersed into a thermal medium, the heavy quark primarily interacts 
with the medium through collisional or radiative processes. 
Both kinds of processes contribute to \emph{transport phenomena}, namely 
they modify the energy and momentum, and thereby position of the heavy quark.
The latter form non-relativistic bound states in QCD with 
$m_h v^2 \ll m_h v \ll m_h$. 
The low-lying ones have most of their support within the Coulomb core of the 
QCD static potential, and thereby constitute the QCD analog of positronium. 
Due to this analogy these bound states are called \emph{quarkonia}. 
The effective field theory ideally suited to describing these bound 
states is potential non-relativistic QCD (pNRQCD), which is obtained by 
successively integrating out the scales $m_h$ and $m_h v \sim \sfrac{1}{r}$ 
(see, e.g., Ref.~\cite{Brambilla:2004jw} for a review). 
When immersed into a thermal medium, the heavy quarkonia primarily interact 
with the medium through dissociative processes leading to thermal broadening, 
or acquire a thermal mass shift that is attributed to a screening of color 
charges. 
Due to the dissociative processes the heavy quark and antiquark may spend 
considerable time in a dissociated state before recombining again. 
Hence, a full dynamical description of in-medium quarkonia must account for 
the transport phenomena that govern the evolution of the dissociated state, too.
The phenomena affecting both types of heavy-quark probes are due to dynamical
processes, and therefore require a real-time formalism to explicitly manifest 
in correlation functions. 

Due to the large value of the heavy-quark mass $m_h$, which is much larger 
than the emergent scale $\Lambda_\text{\tiny{QCD}}$, at which non-perturbative 
effects dominate, the weak-coupling approach is applicable to many analytic 
calculations of properties of heavy-quark systems. 
While the heavy-quark mass $m_h$ is still much larger than the medium 
temperature $T$, in-medium heavy-quark systems show some resilience against 
thermalization. 
Many interactions between the heavy-quark system and its thermal environment 
may be considered as perturbations of vacuum-like properties that are suppressed 
by powers of the ratio of the scales $\sfrac{T}{m_h}$, even for a 
strongly-coupled medium. 
The question, which type of medium interaction or thermal perturbation is 
most relevant strongly depends on the regime determined by the various scales 
of the problem.
The process of quarkonia fully thermalizing with the medium is termed 
\emph{melting} independent of the underlying mechanism. 

The idea of studying the thermal modification of heavy quarkonia in a 
heavy-ion collision is an old one. 
Starting from a confining potential model for the binding of $\mr{J/\Psi}$ 
and the Yukawa-type modification due to the Debye screening of chromoelectric 
fields in a QGP, Matsui and Satz deduced that quarkonia cannot survive 
as bound states once the thermal screening length $1/\md$ becomes as small 
as the average radius $r_\mr{J/\Psi}$~\cite{Matsui:1986dk}. 
They concluded that the modification of the quarkonia rates would be a 
fingerprint of QGP formation. 
Later on, these ideas were extended to the notion of sequential melting, 
namely, that excited quarkonia with weaker binding, and thus larger radii, 
would already melt at lower temperatures, thereby establishing the notion 
that a variety of quarkonia constitutes a thermometer of QGP~\cite{Karsch:2005nk}. 
Melting temperatures of different quarkonia species significantly 
depend in model calculations on the assumption whether the 
underlying potential is of a weak- or strong-binding type, 
where the former would be screened similar to the quark-antiquark 
free energies~\cite{Bazavov:2018wmo, Bazavov:2020teh}, while the latter would retain a robust remnant of 
the confining interaction. 
The former would imply the melting of $\mr{J/\Psi}$ slightly above the 
crossover and of $\mr{\Upsilon(1S)}$ around $T \sim 300\,\mr{MeV}$, while the 
latter might permit $\mr{J/\Psi}$ surviving up to significantly higher 
temperatures and $\mr{\Upsilon(1S)}$ up to 
$T \sim 450\,\mr{MeV}$~\cite{Akiba:2015jwa}. 

Heavy-quark transport has been considered for a similarly long time. 
Early calculations in perturbative QCD predicted large values of the 
heavy-quark diffusion constant $D$~\cite{Svetitsky:1987gq}, while lattice 
calculations in the quenched approximation~\cite{Meyer:2010tt, Banerjee:2011ra, 
Francis:2015daa, Brambilla:2020siz} or heavy-ion phenomenology~\cite{Rapp:2009my} 
seem to prefer a smaller value consistent with a strongly-coupled medium. 
Similarly, since effective field theory calculations~\cite{Laine:2011is} 
indicate in the hadron gas phase a decrease towards the crossover, 
the heavy-quark transport coefficients seem to approach in the crossover 
region global extrema close to the bounds predicted by the AdS/CFT 
correspondence, i.e. $D(2\pi T) \simeq 1$~\cite{Herzog:2006gh}. 
Yet a first-principles calculation that also accounts for dynamical 
quarks in the crossover region is not available yet. 

In heavy-ion collisions heavy quark-antiquark pairs are produced 
in hard processes taking place during the earliest stages. 
If the temperature is sufficiently high, namely $\pi T \gtrsim m_c$, which 
is reached beyond $T \gtrsim 300\,\mr{MeV}$, then charm quark-antiquark pairs 
can be created from the thermal medium as well, and contribute to the bulk 
properties, such as the equation of state. 
Their total number is conserved, since the time scale of their flavor-changing 
electroweak decays exceeds the lifetime of the primordial fireball. 
Hence, heavy-quark, and more specifically, bottom-quark observables probe all stages of the collision, and provide 
almost unique access to its early-time dynamics, and feature prominently in 
the long-term planning of many heavy-ion experiments (see e.g.~\cite{Akiba:2015jwa}). 
Description of in-medium heavy-quark evolution usually relies on perturbative 
QCD, effective field theory or models. 
All of these require some form of non-perturbative input, in particular for 
the strongly-coupled medium at temperatures not too far above the crossover. 
In many cases this input can be provided either by a suitable comparison to 
experimental data or by first-principles lattice gauge theory simulations. 
The most technically challenging aspect of such a lattice computation is that 
it provides Euclidean-time correlators $G(\tau,T)$ that are related to a  
spectral function $\rho(\omega,T)$ encoding the underlying real-time dynamics 
through an integral transform 
(in the following $\beta=\sfrac{1}{T}$), 

\begin{align}
  G(\tau,T) = 
  \int_{-0}^{+\infty} \frac{d\omega}{\omega} K(\omega,\tau,\beta) \rho(\omega,T),
  \label{eq:inverse problem}
\end{align}

where the kernel $K(\omega,\tau,\beta)$ has -- for relativistic quarks -- 
its own temperature dependence:

\begin{align}
  K(\omega,\tau,\beta) = \frac{\cosh \omega\left(\tau-\sfrac{\beta}{2}\right)}{\sinh \sfrac{\omega\beta}{2}}.
  \label{eq:relativistic kernel}
\end{align}

Due to the kernel's symmetry under $\tau \to \beta-\tau$ in 
Eq.~\eqref{eq:relativistic kernel}, the effective number of data in $G(\tau,T)$ 
providing independent information on $\rho(\omega,T)$ is further reduced. 
So far, the inverse problem in Eq.~\eqref{eq:inverse problem} has not been solved 
in any case of interest without including additional assumptions or limitations 
that may have significant impact on the solution.
%~[\ldots]. 

In these proceedings, we review the recent advances in lattice gauge theory 
with respect to the non-pertubative calculation of in-medium heavy-quark 
observables at finite temperature. 
In Section~\ref{sec:transport} we discuss heavy-quark transport, while we 
address heavy quarkiona in Section~\ref{sec:quarkonia}. 
We close with a summary and outlook in Section~\ref{sec:conclusions}.

\section{Heavy-quark transport}
\label{sec:transport}

In the vacuum, open heavy flavors either play a role as hard partonic jets 
at short distances, or in the form of bound states with light degrees of 
freedom at long distances.  
Heavy-light mesons constitute the ground states of such systems. 
In QGP, open heavy flavors may arise either from the eventual melting of 
in-medium heavy quarkonia, or by not binding into a color-singlet state in 
the first place. 
In the following we focus on slowly-moving, non-relativistic heavy quarks, 
for whom collisional medium interactions are dominant. 
The screening of forces between color charges implies that these isolated 
heavy quarks may propagate through the medium without having to form bound 
states, and eventually thermalize with the environment.  
Due to the equipartition theorem, they possess kinetic energy 
$E = \sfrac{\bm{p}^2}{2m_h} \sim T$ and thereby spatial momentum 
$\sqrt{\bm{p}^2} \sim \sqrt{m_h T} \gg T$. 
Successive collisions do not change the momentum $\bm{p}$ significantly and 
can be considered as independent kicks of $\Delta p_i \sim T$; for this 
reason the heavy quarks undergo Langevin-type evolution

\begin{align}
  \frac{d p_i}{d t} = \eta_D p_i + \xi_i(t),
  \quad \braket{\xi_i(t_0) \xi_j(t_1)} = \kappa \delta_{ij} \delta(t_0-t_1),
\end{align}

where the drag coefficient $\eta_D$ and the \emph{heavy-quark momentum 
diffusion coefficient} $\kappa$ are related on general thermodynamic grounds 
by the fluctuation-dissipation theorem as $\eta_D=\sfrac{\kappa}{2m_h T}$. 
The heavy-quark diffusion constant $D = \sfrac{2T^2}{\kappa}$ is related to 
$\kappa$ as well. 
Hence, knowledge of either $\eta_D$, $\kappa$ or $D$ is sufficient to fix the 
equations governing the Langevin dynamics of in-medium heavy quarks 
up to corrections of order $\sfrac{T}{m_h}$~\cite{Bouttefeux:2020ycy}. 

In practice, the path of least resistance is to compute $\kappa$ 
in lattice gauge theory simulations. 
While it could be determined in principle through a Kubo relation 
from the divergence of the transport peak of the spectral function underlying 
the heavy-quark vector correlator at zero frequency, this is notoriously 
hard to compute. 
It was realized that a suitable heavy-quark vector correlator

\begin{align}
  \kappa = 
  \lim\limits_{\omega \to 0} \lim\limits_{m_h \to \infty} 
  \kappa^{(m_h)}(\omega),
  \quad 
  \kappa^{(m_h)}(\omega) =
  \frac{1}{3\chi} \int_{-\infty}^{+\infty} d t e^{\ri \omega t} \int d^3x
  \Braket{\frac{1}{2}\left\{ \mc{F}^i(\bm{x},t),\mc{F}^i(\bm{0},0) \right\} },
\end{align} 

with $\mc{F}^i(\bm{x},t) \equiv m_h \partial_t J^i(\bm{x},t)$,~
$J^i(\bm{x},t)=\bar{\psi}(\bm{x},t) \gamma^i {\psi}(\bm{x},t)$, and 
the quark number susceptibility 
$\chi =\beta \int d^3x
\Braket{\frac{1}{2}\left\{ J^0(\bm{x},t),J^0(\bm{0},0) \right\} }$,
reduces -- after integrating out the heavy-quark fields -- to a 
chromoelectric field-strength correlator in Euclidean time~\cite{Caron-Huot:2009ncn}

\begin{align}
  G_{E,R}(\tau,T) =
  -\frac{1}{3} \sum_{i=1}^3 \frac{\braket{\mathrm{Re\,Tr} U(\beta,\tau) gE_i(\tau,\bm{0})U(\tau,0)gE_i(0,\bm{0})}}{\braket{\mathrm{Re\,Tr} U(\beta,0)}},
  \label{eq:chromoelectric correlator}
\end{align}

which is amenable to the lattice evaluation. 
$P=\braket{\mathrm{Re\,Tr} U(\beta,0)}$ is the trace of the Polyakov loop. 
Since such gluonic operators are very noisy quantities, efficient 
noise suppression techniques are indispensable for their calculation. 
Employing multi-level algorithms in the quenched approximation 
$G_{E,R}(\tau,T)$ has been obtained on the lattice by various groups in the 
past~\cite{Meyer:2010tt, Banerjee:2011ra, Francis:2015daa, Brambilla:2020siz}. 
As this correlator requires multiplicative renormalization, which 
is to date known at the one-loop level~\cite{Christensen:2016wdo}, 
estimates of the renormalization factor had been obtained earlier 
using tadpole factors. 
After obtaining the continuum limit of the chromoelectric correlator 
in Eq.~\eqref{eq:chromoelectric correlator}, it can be related through 
the kernel $K(\omega,\tau,\beta)$ in Eq.~\eqref{eq:relativistic kernel} 
to the spectral function via 

\begin{align}
  G_{E,R}(\tau,T) = 
  \int_{-0}^{+\infty} \frac{d\omega}{\pi} K(\omega,\tau,\beta) \rho(\omega,T),
  \quad 
  \kappa = \lim\limits_{\omega \to 0} \frac{2T \rho(\omega,T)}{\omega}.
  \label{eq:hq spectral function}
\end{align}

In practice, the inverse problem in Eq.~\eqref{eq:hq spectral function} 
is solved using models for the spectral function that enforce the known 
limiting behaviors in the IR and in the UV. 

Novel lattice calculations have improved this situation 
considerably. 
The key ingredient to these new approaches is the use of the gradient 
flow instead of the multi-level algorithm for the noise reduction. 
The Yang-Mills gradient flow evolves the gauge fields along a 
fictitious fifth direction termed \emph{flow time} $\tau_F$~\cite{Luscher:2010iy}. 
The flowed fields are defined via 

\begin{align}
  \frac{\partial V_\mu(x,\tau_F)}{\partial \tau_F} 
  & = D_\nu G_{\nu\mu},
  \quad 
  V_\mu(x,\tau_F=0) \equiv U_\mu(x), \\
  D_\mu V_\nu(x,\tau_F) 
  & \equiv \partial_\mu + [V_\mu(x,\tau_F),V_\nu(x,\tau_F)], \\ 
  G_{\nu\mu}(x,\tau_F) 
  & \equiv 
  \partial_\mu V_\nu(x,\tau_F) -
  \partial_\nu V_\mu(x,\tau_F) +
  [V_\mu(x,\tau_F),V_\nu(x,\tau_F)],
  \label{eq: flow action}
\end{align}

where the discretization in Eq.~\eqref{eq: flow action} does not 
have to coincide with the discretization used in the action. 
Any composite local operator computed from the fields $V_\mu(x,\tau_F)$ at 
large enough finite flow time yields in the zero flow-time limit a linear 
combination of renormalized operators within the same representation, 

\begin{align}
  \lim\limits_{\tau_F \to 0} O[V_\mu(x,\tau_F)] = 
  \sum\limits_i c_i(\tau_F) O_i^\mathrm{R}(x).
  \label{eq:flowed operator}
\end{align}

Hence, in practice the chromoelectric correlator defined in 
Eq.~\eqref{eq:chromoelectric correlator} is computed using 
fields $V_\mu(x,\tau_F)$ at a wide enough range of flow times $\tau_F$ 
for different lattice spacings. 
Then $O[V_\mu(x,\tau_F)]$ is extrapolated to the continuum limit 
at each value $\tau_F$ of fixed flow time. 
These continuum results are extrapolated to zero flow time in 
an ensuing step. 
While the functional form of the continuum extrapolation is well 
motivated by the Symanzik effective theory, the details of the 
zero flow-time extrapolation are often somewhat ad hoc. 
Nonetheless, since varying the choice of the flow action in 
Eq.~\eqref{eq: flow action} permits changing the details of the 
finite flow-time artifacts and change the slope (and higher 
order $\tau_F$ dependence), these latter extrapolations appear 
to be usually well under control. 

If the initial operator was chosen such that it cannot mix with 
other operators, the renormalized continuum result is obtained. 
While the comparison of the renormalized continuum results obtained 
via gradient flow, or via multi-level algorithm with one-loop 
renormalization factor suggests small differences between the 
renormalization factors, the overall time and temperature dependence 
of these correlators is in good agreement given the quoted 
errors~\cite{Altenkort:2020fgs, Mayer-Steudte:2021hei}. 
Obviously, the need for modeling the spectral functions to solve 
the inverse problem in Eq.~\eqref{eq:hq spectral function} is 
not significantly affected by the change of the lattice setup used 
to obtain the Euclidean correlator in the continuum limit. 
The most recent results for $\kappa$ are quantitatively consistent, albeit still with substantial uncertainties. 

The use of gradient flow for computing the heavy-quark transport 
coefficients opens up two new windows of opportunity. 
The first and most obvious one is that the gradient flow as a means 
for noise reduction is applicable in full QCD as well, while the 
multi-level algorithm is not due to the non-local nature of the 
fermion determinant. 
Hence, we may expect the first calculations of the heavy-quark momentum 
diffusion coefficient $\kappa$ in full QCD in the near future. 
The second and more subtle one is that the gradient flow as a 
means for renormalization permits to compute full chromoelectric 
or chromomagnetic correlation functions, such as 

\begin{align}
  G_B(\tau,T) &= \int d^3x
  \frac{\Braket{\tr{U(\beta;\tau) B_i(\tau) U(\tau;0)B_i(0)}}}{3P},
  \label{eq:full chromomagnetic correlator} \\
  G_E(\tau,T) &= \int d^3x
  \frac{\Braket{\tr{U(\beta;\tau) E_i(\tau) U(\tau;0)E_i(0)}}}{3P},
  \label{eq:full chromoelectric correlator}
\end{align}

for which the renormalization with other means is still an 
unsolved problem. 
The real part of the chromomagnetic correlator in 
Eq.~\eqref{eq:full chromomagnetic correlator} is related to the 
$\mathcal{O}(\sfrac{T}{m_h})$ corrections to $\kappa$. 
Preliminary results of the first such calculations have been 
reported at this conference~\cite{Altenkort:2021ntw, Mayer-Steudte:2021hei}. 
The imaginary part in Eq.~\eqref{eq:full chromoelectric correlator} 
is related to $\gamma$, the dispersive counterpart of $\kappa$, which is 
related to the thermal mass shift of quarkonia in pNRQCD~\cite{Brambilla:2019tpt}.

\section{Heavy quarkonia}
\label{sec:quarkonia}

\subsection{General features}
\label{sec:general}

In the vacuum, quarkonia are among the most simple bound state 
problems of QCD. 
The heavy-quark mass $m_h$ is by far the largest scale in the problem. 
By integrating out $m_h$ (in terms of a Foldy-Wouthuysen transform 
similar to the textbook-level one used in the relativistic treatment 
of hydrogen) one arrives at the effective field theory of 
non-relativistic QCD (NRQCD)~\cite{Manohar:1983md}, which is valid 
at the soft scale of typical momenta $p \sim m_h v \ll m_h$. 
The heavy-quark or antiquark fields are described as two separate 
2-component Pauli spinors, while heavy quark-antiquark pair creation and 
annihilation is realized through four-fermion interactions suppressed by 
the ratio of the scales, i.e. the heavy-quark velocity $v=\sfrac{p}{m_h}$, 
which is of the order of the strong coupling constant $v \sim \as(m_h)$. 

For a heavy quark-antiquark pair, it is possible to further simplify the 
problem. 
Since their low-lying bound states are compact objects, a multipole expansion 
in the relative coordinate between quark and antiquark can be applied. 
Namely, $\sfrac{r}{R}$ is a small expansion parameter, with $\bm{r}$ being 
the relative coordinate and $\bm{R}$ being the center-of-mass coordinate. 
For a compact bound state mostly dominated by the Coulomb core of the QCD 
interaction, as predicated by asymptotic freedom, the inverse relative 
coordinate is a scale of the order of typical relative momenta 
$\sfrac{1}{r} \sim p \sim m_h v$, which is integrated out by means of the 
multipole expansion. 
One arrives at the effective field theory called potential non-relativistic 
QCD (pNRQCD)~\cite{Brambilla:1999xf}, whose Wilson coefficients depend on 
the relative coordinate $\bm{r}$ (see, e.g., Ref.~\cite{Brambilla:2004jw} 
for a review). 
The dynamical degrees of freedom of pNRQCD at the ultra-soft scale 
$E \sim \sfrac{\as}{r} \sim m_h v^2$ are bosonic color-singlet or 
-octet fields, $S$ or $O^a$, as well as light degrees of freedom. 
The color-singlet and -octet fields explicitly couple to the ultra-soft 
chromoelectric fields through dipole interactions 
$\propto \bm{r} \cdot \bm {E}^a$ that are suppressed 
by one power of the relative coordinate. 
Some of the non-local Wilson coefficients play the role of the different 
contributions to the heavy quark-antiquark potential in QCD. 
However, there are non-potential contributions (i.e. terms that couple 
singlet- and octet-fields or different octet-fields among each other) 
that contribute as well in pNRQCD beyond the tree-level.  

At leading order (LO), the attractive potential $V_s = -C_F\sfrac{\as}{r}$ 
of the singlet field $S$ in pNRQCD is the \emph{static potential} of QCD. 
Similarly, at this order the repulsive potential 
$V_o = +\sfrac{1}{6}~\sfrac{\as}{r}$ of the octet field $O^a$ is Coulombic, 
too. 
The singlet-singlet correlator in pNRQCD at leading order  can be related to 
the Wilson loop in QCD by matching

\begin{align}
  W(r,t) =
  \braket{ \exp{\ri g \oint_{r \times t} dz^\mu A_\mu} }_\mr{QCD}
  = 
  \braket{ S(r,0) S^\dagger(r,t) }_\mr{pNRQCD}^{LO}. 
\end{align}

It corresponds to quarkonia made from infinitely heavy constituents. 
In the vacuum, the large time behavior of the Wilson loop is given 
in terms of the ground state, the QCD static energy $E(r)$, 

\begin{align}
  E(r) = -\ri \lim\limits_{t \to \infty} 
  \frac{\partial W(r,t)}{\partial t}, 
  \label{eq:vacuum static energy}
\end{align}

which coincides with $V_s$ at small enough distances. 
Such a stable ground state is represented by a distinct and well-separated 
delta function in the spectral function and would manifest as a plateau in 
the effective mass at sufficiently large imaginary time.

However, an unstable, quasiparticle excitation would feature 
locally in the spectral function as a regularized, (skewed) Breit-Wigner 
peak that is distinct and well-separated from other spectral features. 
Its skewing is due to the non-trivially time-dependent UV part 
of the correlator~\cite{Burnier:2012az}. 
The tails of this Breit-Wigner peak would have to be regularized through 
the interplay between different states, or non-potential interactions 
with the medium that are not known in sufficient detail. 
While such a quasiparticle state cannot lead to a plateau in the effective 
mass at imaginary 
time (limited to $\tau < \beta$), 
the origin of the absence of a plateau in the effective mass is not 
immediately obvious. 
First, such absence could be caused even with a stable ground state by a 
time direction that is simply too short to permit decay of all excited states.
Second, such absence could be caused by some quasiparticle remnant of the 
$T=0$ ground state acquiring a finite width at $T>0$, which cannot lead to 
a plateau as stated,  while the right hand side in 
Eq.~\eqref{eq:vacuum static energy} might still become time independent. 
Third, such absence could be caused by a complete \emph{melting} in the 
sense that there is no distinct quasiparticle remnant of the $T=0$ ground 
state anymore. 
In the latter case, the right hand side in Eq.~\eqref{eq:vacuum static energy} 
would cease to be time-independent even in real time. 
Similar considerations apply with straightforward adaptations to 
finite mass quarkonia in a non-relativistic or relativistic formulation. 

\subsection{Weak-coupling approach}

\begin{figure}
\center
\includegraphics[width=10cm]{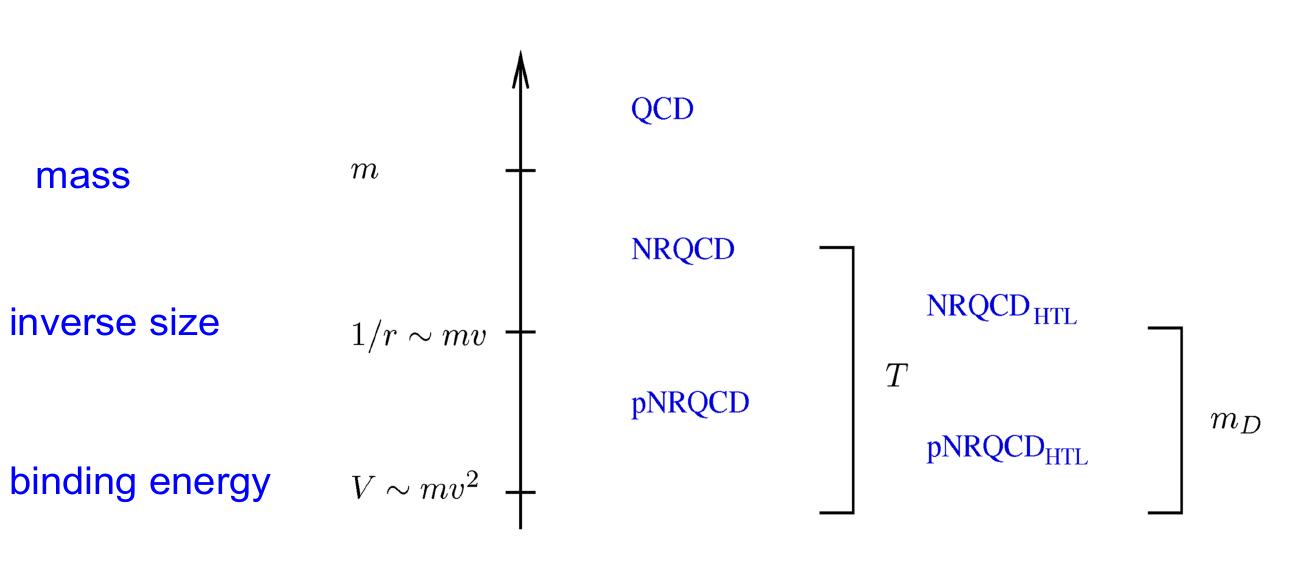}
\caption{Relevant scale hierarchies for in-medium quarkonia.}
\label{fig:scales}
\end{figure}
In the strict weak-coupling approach, the non-relativistic hierarchy 
$m_h \gg m_h v \gg m_h v^2$ is accompanied by a thermal hierarchy 
$\pi T \gg \md \gg g^2 T$, where the Debye mass $\md \sim gT$ is 
related to the scale of chromoelectric fields and $g^2 T$ to the 
scale of chromomagnetic fields. 
In both instances, the thermal coupling $g$ is understood as a coupling 
at one of the two higher thermal scales. 

Since the thermal coupling $g$ or rather the ratio 
$\sfrac{\md}{\pi T} \sim \sfrac{g}{\pi}$ is still large at temperatures far 
above the QCD transition, the convergence of perturbation theory for thermal 
observables is in many cases not obvious or even very poor in practice. 
Moreover, the magnetic scale $g^2T$ is inherently non-perturbative even 
at the highest temperatures, since the chromomagnetic fields are still 
subject to a confining three-dimensional $\mathrm{SU}(3)$ pure gauge theory. 
For these reasons, practical weak-coupling calculations are hampered in 
finite temperature QCD by the emergence of non-perturbative contributions 
from the chromomagnetic scale $g^2T$ at some order, and by the need to 
resum contributions from the scale of the Debye mass. 

Physically distinct regimes emerge depending on the relative ordering between the non-relativistic and thermal scales. 
The QCD static potential at finite temperature has been computed 
at leading order using the Hard Thermal Loop (HTL) approach. 
This calculation~\cite{Laine:2006ns} identified a complex potential in the 
chromoelectric screening regime $\sfrac{1}{r} \sim \md \ll \pi T$,

\begin{align}
  V_s^\mr{HTL}(r,T) = -C_F \as 
  \left\{ \frac{e^{-r\md}}{r} + \md + \ri T \phi(r\md) \right\}, 
  \phi(x) = 2\int\limits_{0}^{\infty} \frac{dz~z}{(z^2+1)^2} 
  \left\{ 1-\frac{\sin(zx)}{zx} \right\},
  \label{eq:Laine potential}
\end{align}

whose real part is Debye screened and coincides in this regime with the 
gauge-dependent singlet free energy (in Coulomb gauge) 
$F_S^\mr{HTL}(r,T) = -C_F \as \left\{ \frac{e^{-r\md}}{r} + \md \right\}$ up to 
the order $\mc{O}(g^4)$. 
Yet this result contains a non-trivial imaginary part as well, 
$\mr{Im}~V_s^\mr{HTL}(r,T) = \mc{O}(g^2 T)$, that would vanish for $r \to 0$.  
Shortly after, this result was confirmed in a separate 
calculation~\cite{Brambilla:2008cx} that also considered the short distance 
regime $\Delta V \sim \sfrac{\as}{r} \ll \sfrac{1}{r} \ll \md \ll \pi T$; in the latter 
regime the real part is vacuum-like without hints of screening, 

\begin{align}
  V_s^\mr{pNRQCD}(r,T) = - \frac{C_F \as}{r} + r^2 T^3 \left\{ \mc{O}(g^4) 
  + \ri \mc{O}\left( g^4, \frac{g^6}{(rT)^2} \right) \right\},
  \label{eq:BGPV potential}
\end{align}

while the imaginary part retains a non-zero value even as $r \to 0$. 

As a consequence of these developments a new picture of quarkonia melting 
has begun to emerge. 
Namely, the presence of an imaginary part of the potential, which can be 
understood as being due to Landau damping and singlet-to-octet transitions, 
would lead to a dissociaton through decorrelation. 
Therefore, a dynamical picture of melting would replace the static picture 
suggested by Matsui and Satz~\cite{Matsui:1986dk}. 
Nevertheless, these weak-coupling results are unable to address questions 
concerning the relative importance of these different effects at 
phenomenologically interesting temperatures, i.e. in the broad crossover 
region and somewhat above, where the underlying assumption of a strictly 
weakly-coupled thermal scale hierarchy might be too optimistic. 

\subsection{Heavy quarkonia with relativistic quarks}
\label{sec:relativistic}

One possible approach to tackle the question of heavy-quarkonia melting 
non-perturbativly -- albeit in a qualitative manner -- has been to study 
screening correlators on the 
lattice~\cite{Karsch:2012na, Bazavov:2014cta, Bazavov:2019www}. 
When studying temporal correlation functions in finite temperature lattice 
simulations one is generally hampered by having a very small number 
of data along the necessarily very short Euclidean time direction, since 
$a N_\tau = \beta$. 
Moreover, the matter is made worse by the symmetry of the relativistic 
correlator and kernel under $\tau \to \beta-\tau$, see 
Eq.~\eqref{eq:relativistic kernel}, effectively reducing the amount 
of independent information by a factor two. 
As a consequence there are simply insufficient data for resolving subtle, 
temperature-induced changes of such a temporal correlator with lattices 
of the foreseeable future. 
Moreover, since the kernel $K(\omega,\tau,\beta)$ for relativistic quarks 
in Eq.~\eqref{eq:relativistic kernel} has explicit temperature dependence, 
the temperature dependence of the correlator $G(\tau,T)$ does not originate 
only in the temperature dependence of the spectral function $\rho(\omega,T)$. 

Considering instead a spatial or screening correlator has two advantages. 
First, the spatial extent of the lattice may be much larger than the inverse 
temperature -- for a given lattice spacing $a$ the number $N_\sigma$ of 
data effectively decouples from $N_\tau$. 
Second, the corresponding kernel has no explicit temperature dependence. 
Yet the price to pay is a more complicated relation between the screening 
correlator $G(z,T)$ (here in the $z$ direction) and the underlying 
finite momentum spectral function $\rho(\omega,p_z,T)$, namely, 

\begin{align}
  G(z,T) 
  &= \int\limits_0^{\beta} d \tau \int\limits d^2x_\perp
\Braket{ J(\tau,\bm{x}_\perp,z) J^\dagger(0) } 
  = \int\limits_{0}^{\infty} \frac{2 d \omega}{\omega} \int\limits_{-\infty}^{\infty} d p_z e^{i p_z z} \rho(\omega,p_z,T)
  \stackrel{z \to \infty}{\sim} e^{-M(T)z}.
  \label{eq:screening correlator}
\end{align}

The exponential decay of a Euclidean quark-antiquark correlation function 
at large separations is independent of the direction between source and 
sink at zero temperature, or more generally speaking, whenever a stable 
mesonic ground state or a quasiparticle one with mostly unmodified 
properties exists. 
The effects of the different boundary conditions (periodic in space, 
antiperiodic in time) cancel as long as there is a stable 
meson~\cite{Bazavov:2014cta}. 
Yet if there is no stable meson, but a non-interacting quark-antiquark pair, 
the screening mass is of the form 
$M(T) = \sum_i \sqrt{m_{h,i}^2+(\pi T)^2}$ 
due to the anti-periodic boundary condition in time. 
When the ground state melts, the screening mass smoothly changes from its 
temperature-independent vacuum value $M_\mr{vac}$($=M_\mr{PDG}$) to 
the asymptotic, temperature-dependent behavior of the non-interacting 
quark-antiquark pair. 

\begin{figure}
\center
\includegraphics[width=6cm]{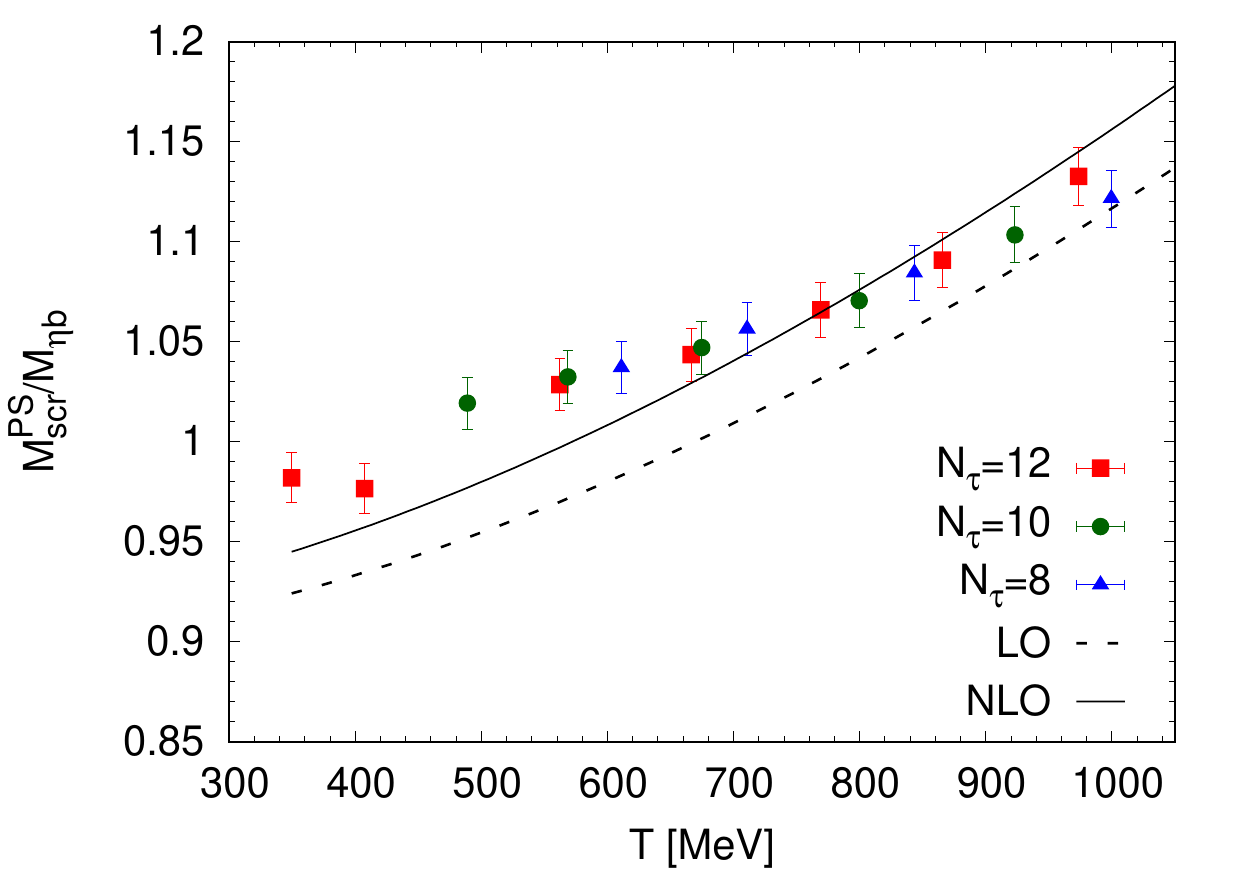}
\includegraphics[width=6cm]{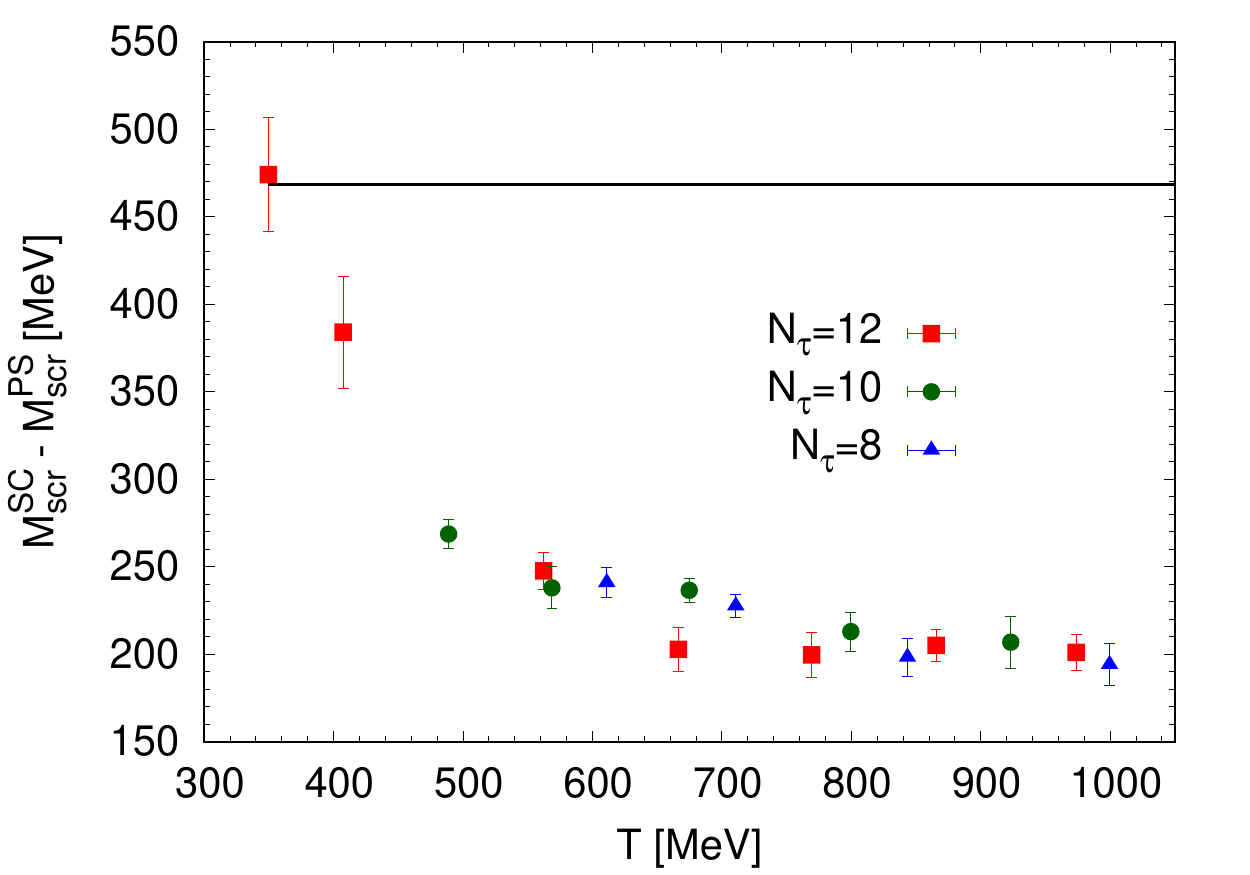}
\caption{Masses of $T>0$ bottomonia screening 
correlators using HISQ action in (2+1)-flavor QCD. 
The ratio $\sfrac{M(T)}{M(0)}$ in the pseudocsalar channel is consistent 
with 1 up to $T \approx 400\,\mr{MeV}$ (left). 
The rapid drop of the difference of scalar and pseudoscalar masses 
from $T \approx 350\,\mr{MeV}$ onward indicates the melting of 
$\chi_{b0}$, but tampers out in a slow decrease due to heavy-quark 
mass effects (right). 
Figures from Ref.~\cite{Petreczky:2021zmz}.}
\label{fig:bottomonium screening masses}
\end{figure}

Following these considerations, bottomonia screening correlators have 
been computed using the highly improved staggered quark (HISQ) action 
combined with the tree-level improved Symanzik gauge action on (2+1)-flavor 
QCD lattices close to the physical point ($m_l=\sfrac{m_s}{20}$, and a 
physical strange quark)~\cite{Petreczky:2021zmz} in an extension of an 
earlier study in the same HotQCD setup that had been focusing on open and 
hidden charm screening correlators~\cite{Bazavov:2014cta}. 
This earlier study found that the most compact ground state charmonia 
($\mr{J/\Psi}$ and $\eta_c$) melt at temperatures higher than $T \gtrsim 200\,\mr{MeV}$, 
while the more spatially extended $\chi_{c}$ states 
melt already slightly above the crossover. 
Hence, any charmonia have already been fully melted before they begin to 
contribute significantly to the equation of state beyond 
$T \gtrsim 300\,\mr{MeV}$ as reported at this 
conference~\cite{Weber:2021hro} for the HISQ action. 
The finite temperature lattices in the new study~\cite{Petreczky:2021zmz} 
had $N_\tau=12,~10$, and $8$, covering a temperature window of 
$T \sim [350,1000]\,\mr{MeV}$. 
For the heavy-quark currents in Eq.~\eqref{eq:screening correlator} only 
local interpolating operators and point sources had been employed. 

The main conclusions of this study are indicated in 
Fig.~\ref{fig:bottomonium screening masses}, and were reported together with 
many technical details at this conference~\cite{Petreczky:2021hyd}. 
Namely, the most compact ground state bottomonia ($\mr{\Upsilon(1S)}$ and $\eta_b$) 
are mostly unmodified until at least $T \gtrsim 400\,\mr{MeV}$, while the 
lowest, larger bottomonia ($\chi_{b0}$, $h_b$) begin the melting already after $T \gtrsim 350\,\mr{MeV}$. 
The qualitative modifications compared to charmonia screening 
correlators can be quantitatively understood in an effective field 
theory framework in terms of higher order corrections, since the onset 
of the naively expected high temperature behavior is suppressed by 
powers of $\sfrac{T}{m_b}$. 
Moreover, the staggered discretization artifacts were found to be 
commensurate with the size of other errors.

This program of studying screening correlators for $N_f=3$ has been extended 
towards the electroweak scale using $\mc{O}(a)$-improved Wilson fermions 
in the chiral limit. 
Due to mild cutoff effects precise continuum results could be obtained. 
These showed that deviations from the leading order -- while being only 
a few percent down to $T \sim 1\,\mr{GeV}$ -- could not be captured 
quantitatively by the next-to-leading order even at $T \sim 80\,\mr{GeV}$, 
and were reported at this conference~\cite{lat2021laudicina}. 
Of course, these findings apply to heavy quarkonia as well at such 
high temperatures. 

\subsection{Heavy quarkonia using NRQCD}
\label{sec:NRQCD}

In general, the velocity $v$ plays the role of an expansion parameter in 
NRQCD, which has been formulated on the lattice~\cite{Thacker:1990bm, 
Lepage:1992tx}, too. 
The computation of heavy-quark propagators simplifies to the solution of 
an initial value problem, since there is no backward propagating component. 
Suppressed higher order operator insertions as corrections to the identity 
alternate with propagation by one time step using the leading-order operator. 
This must be contrasted with the relativistic treatment, where obtaining 
a heavy-quark propagator requires a matrix inversion. 
Yet lattice NRQCD has one major drawback compared to a relativistic 
treatment or a treatment in the fully static limit. 
Namely, since it requires $m_h \gtrsim \sfrac{1}{a}$, the continuum limit 
cannot be taken.

With regard to thermal correlators, these properties of NRQCD have two 
main consequences. 
First, for any given heavy-quark mass $m_h$ and any $N_\tau$, there is a 
rather strong upper limit on the temperatures to which it can be applied. 
This needs to be contrasted to relativistic formulations of heavy quarks, 
where there is a rather strong lower limit on the temperatures for which 
cutoff effects (parameterized via an expansion in $a m_h$) are 
sufficiently small. 
Second, the inverse problem of reconstructing the spectral function 
from a thermal correlator such as the one in Eq.~\eqref{eq:inverse problem} 
is alleviated for any non-relativistic formulation. 
Any such formulation leads to a simplified, temperature-independent integral 
kernel, $e^{-\omega \tau}$, 
which is simply the kernel of a Laplace transform, and breaks the symmetry 
under $\tau \to \beta-\tau$, and thus effectively doubles the amount of 
independent information compared to the relativistic case, 
see Section~\ref{sec:relativistic}. 

In-medium NRQCD bottomonia have been studied for a long time by the FASTSUM 
collaboration~\cite{Aarts:2014cda} and through the well-established 
HotQCD setup~\cite{Kim:2018yhk, Larsen:2019bwy, Larsen:2019zqv, 
Larsen:2020rjk}. 
One central difference in the approaches of the two groups has been 
the temperature scan in a fixed scale approach (through variation of 
$N_\tau$) of FASTSUM (most recent results on gen. 2L anisotropic lattices 
with (2+1) flavors of Wilson fermions, $m_\pi=236\,\mr{MeV}$, $\sfrac{a_s}{a_t}=3.45$)
or through variation of the lattice spacing $a$ in the HotQCD setup 
(HISQ/Tree action with $m_l=\sfrac{m_s}{20}$, see 
Section~\ref{sec:relativistic}). 
A second key difference in the most recent studies is that FASTSUM has 
focused so far on the use of point sources for the NRQCD correlators, 
while HotQCD has established the use of extended sources.

\begin{figure}
\center
\includegraphics[width=6cm]{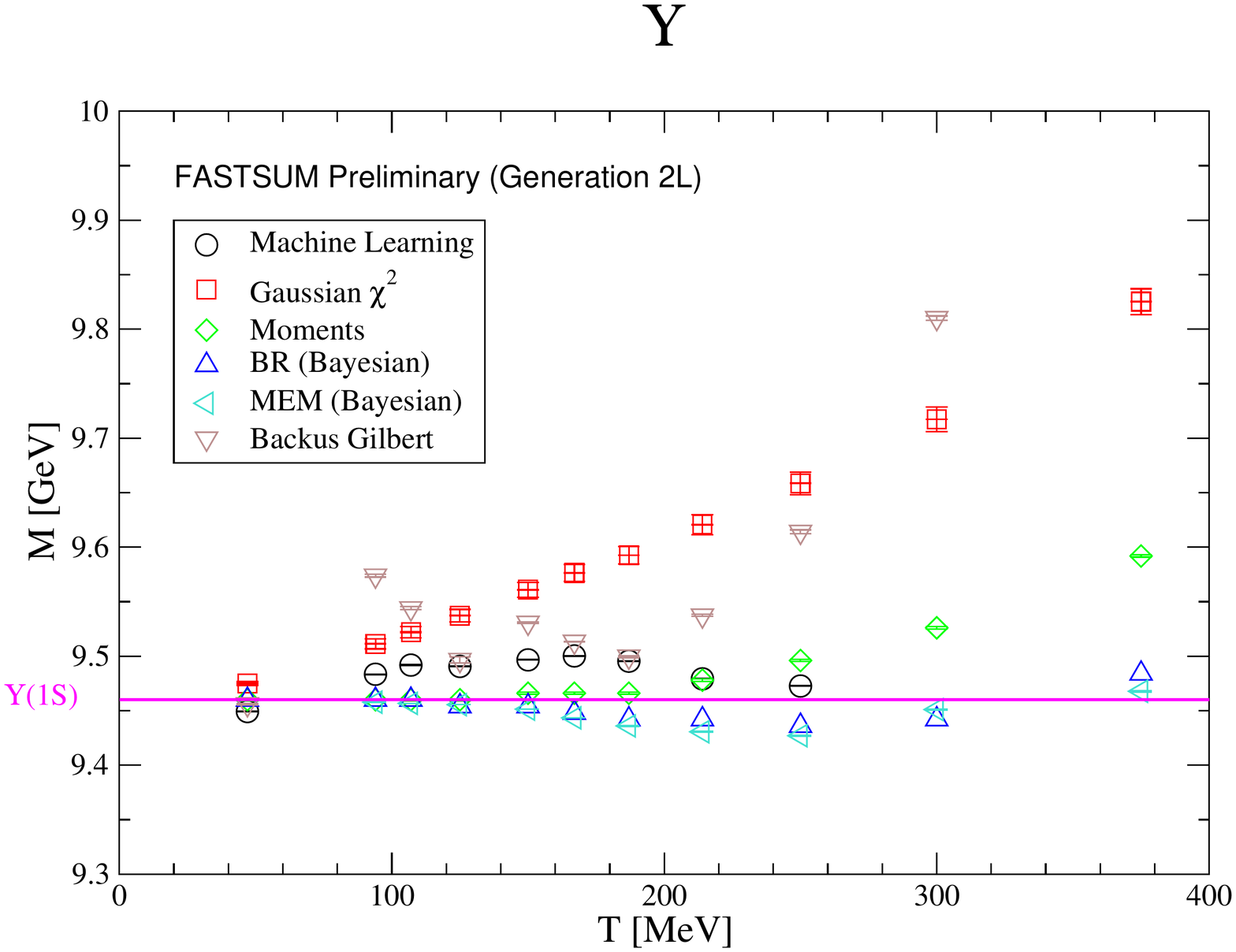}
\includegraphics[width=6cm]{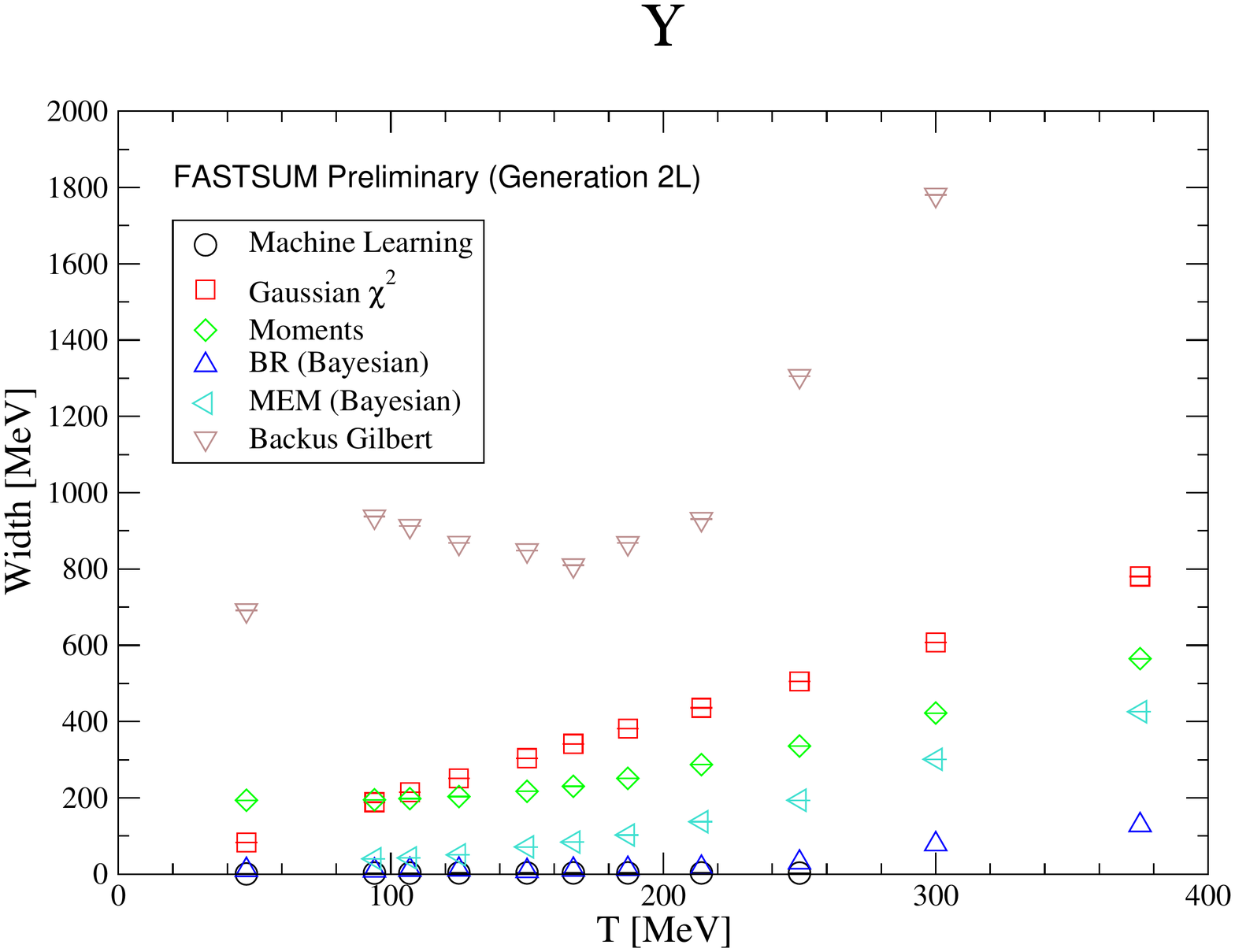}
\caption{Upsilon correlators using NRQCD on gen. 2L FASTSUM 
lattices in (2+1)-flavor QCD. 
Different analysis methods may suggest opposite signs for the 
thermal mass shift (left). 
All analysis methods eventually find a finite width (monotonically 
increasing with temperature except for Backus-Gilbert method), although 
for some consistent with zero up to $T \lesssim 250\,\mr{MeV}$ (right). 
Figures courtesy of C.~Allton.}
\label{fig:bottomonium NRQCD masses}
\end{figure}

The FASTSUM correlators obtained on the gen. 2L anisotropic lattices 
have been analyzed using a wide range of techniques -- plain Gaussian 
$\chi^2$ minimization, moments of the correlator, machine learning via 
Kernel Ridge Regression~, the Backus-Gilbert method, and the two 
Bayesian approaches of Maximum Entropy Method (MEM) or 
Bayesian Reconstruction (BR). 
So far, a clear picture has not emerged yet; 
applied to the same correlators, the mass shifts may be positive 
or negative, and the widths do not reveal a pattern consistent 
between the various methods, as indicated in Fig.~\ref{fig:bottomonium NRQCD masses}; 
preliminary results were discussed at this conference~\cite{Spriggs:2021jsh, Page:2021ohe, Offler:2021fmg} 
and summarized in further conference proceedings~\cite{Spriggs:2021dsb}.

The HotQCD results paint quite a different picture. 
An earlier analysis with point sources concluded that a plateau is 
not even reached in the pseudoscalar channel at $T=0$ 
on fine lattices~\cite{Kim:2018yhk}. 
After accounting for this observation negative finite temperature 
mass shifts were identified. 
In order to overcome the lack of convergence to the ground state, 
they turned to Gaussian smeared sources that achieve a substantial 
suppression of the UV part of the correlator~\cite{Larsen:2019bwy}. 
While extended sources distort the UV part, it was found that the 
distortion is very similar at $T=0$ and at finite temperature. 
Therefore, as a key component of the HotQCD analysis the $T=0$ 
UV part was subtracted from finite temperature correlators. 
Analyzing the $T=0$ correlator with multi-exponential fits and 
subtracting the ground state contribution, the remnant was considered 
as the full UV part, and subtracted from the finite temperature correlator.  
In earlier calculations~\cite{Petreczky:2017aiz} a similar procedure had 
been already applied in the static limit, i.e. to Wilson loops.
This UV subtraction produces an effective mass that is almost linear 
in $\tau$ at small times, namely $\tau \lesssim \sfrac{\beta}{2}$, 
but falls rapidly at $\tau \gtrsim \sfrac{\beta}{2}$. 

On very general grounds, one may expect for any finite temperature 
spectral function $\rho_X(\omega,T)$ corresponding to a correlator with 
spatially extended interpolating fields that it exhibits three distinct features. 
First -- unless it has reached the stage of complete melting -- a well-separated, thermally-modified lowest spectral feature $\Omega_X(T)$, which could be traced back to the original $T=0$ ground state. 
Second, a UV part due to excited states of the quark-antiquark pair immersed in the QCD medium. 
And third, a \emph{low-energy tail} that is caused by 
excitations of the QCD medium itself that propagate backwards in 
imaginary time~\cite{Bala:2021fkm}.

\begin{align}
  \rho_X(\omega,T) 
  = \rho_X^{\Omega}(\omega,T)
  + \rho_X^\mr{UV}(\omega,T) 
  + \rho_X^\mr{tail}(\omega,T).
  \label{eq:spectral decomposition}
\end{align}

These general considerations apply to any sufficiently extended 
operator, whether this is realized via NRQCD with smeared sources, 
or -- as we will peruse later -- Wilson loops. 
Since bottomonia vector correlators have a splitting of 
$m_{\mr{\Upsilon(2S)}}-m_{\mr{\Upsilon(1S)}} \sim 600\,\mr{MeV}$, 
one may assume sufficient separation at least for $\mr{\Upsilon(1S)}$ 
at low temperatures. 
The UV subtracted correlators correspond to 
$\rho_X^\mr{UV}(\omega,T) -\rho_X^\mr{UV}(\omega,0) \approx 0$, 
while rapid falloff at large $\tau \gtrsim \sfrac{\beta}{2}$ 
originates in $\rho_X^\mr{tail}(\omega,T)$. 
Taking these features into account, the authors analyzed the 
subtracted correlators through fits of the form 

\begin{align}
  G^\mr{sub}(\tau,T) 
  &= A^{\Omega}(T) e^{-\Omega(T)\tau+(\Gamma^G(T))^2 \frac{\tau^2}{2}} 
  + A^\mr{tail}(T) e^{-\omega^\mr{tail}(T)\tau},\quad
  \omega^\mr{tail}(T) \ll \Omega(T).
  \label{eq:gaussian+tail}
\end{align}

While Eq.~\eqref{eq:gaussian+tail} does not correspond to the 
Breit-Wigner shape that is necessary in the quasiparticle picture, 
it can be understood as a regularization of a Breit-Wigner peak 
through a Gaussian. 
The width $\Gamma^G(T)$ of the Gaussian then has to be rescaled 
to the corresponding effective width 
$\Gamma(T)=\sqrt{2\ln2}\Gamma^G(T)$ of the Breit-Wigner that it 
imitates. 
Since the extended correlators do not provide sufficient number 
of independent data over a wide enough range of $\tau$, further 
features such as skewing (cf. Section~\ref{sec:general}) could 
not be resolved so far. 
Moreover, for the same reason the low-energy tail was modeled 
as a delta function. 
This analysis found a thermal mass shift for the $\mr{\Upsilon(1S)}$ that 
is almost negligible, while its thermal width increases monotonically 
with the temperature~\cite{Larsen:2019bwy}. 

With approximate eigenfunctions of a Cornell potential used as 
interpolating fields when solving a GEVP, similar findings were 
obtained even for the first few excited states~\cite{Larsen:2019zqv}, 
with similarly small mass shifts, but significantly larger widths. 
The authors went one step further, and solved the Schr\"odinger 
equation for the $T=0$ lattice NBS amplitudes 

\begin{align}
  \left\{ -\frac{\Delta}{2m_b} + V(r) \right\} 
  \phi_{\mr{\Upsilon(nS)}} = E_{\mr{\Upsilon(nS)}} \phi_{\mr{\Upsilon(nS)}}
  \label{eq:Schrodinger}
\end{align}

using two different energy levels to determine the effective 
bottom quark mass and reconstruct the underlying potential. 
This potential was found to be quantitatively consistent with 
the one obtained from Wilson loops. 
Extending the approach of NBS amplitudes to finite temperature, 
the authors found minimal temperature dependence of the BS 
amplitudes $\phi_{\mr{\Upsilon(nS)}}$ at small $\tau$, but a substantial 
$\tau$ dependence for the NBS amplitudes of the excited states~\cite{Larsen:2020rjk}. 

The NBS amplitudes $\phi_{\mr{\Upsilon(nS)}}$ of Ref.~\cite{Larsen:2020rjk} were 
processed in a deep neural network (DNN)~\cite{Shi:2021qri} solving 
Eq.~\eqref{eq:Schrodinger} to address the question whether the specific 
fit form in Eq.~\eqref{eq:gaussian+tail} may have led to substantial 
bias in Refs.~\cite{Larsen:2019bwy, Larsen:2019zqv}. 
The DNN results positively confirmed the earlier calculation. 
Another fully independent calculation of the quark-antiquark potential 
by the FASTSUM collaboration employing lattice NBS amplitudes when solving 
Eq.~\eqref{eq:Schrodinger} was reported 
at this conference~\cite{Spriggs:2021ieo}, too.

\subsection{Heavy quarkonia using static quarks}
\label{sec:static}

\begin{figure}
\center
\includegraphics[width=6cm]{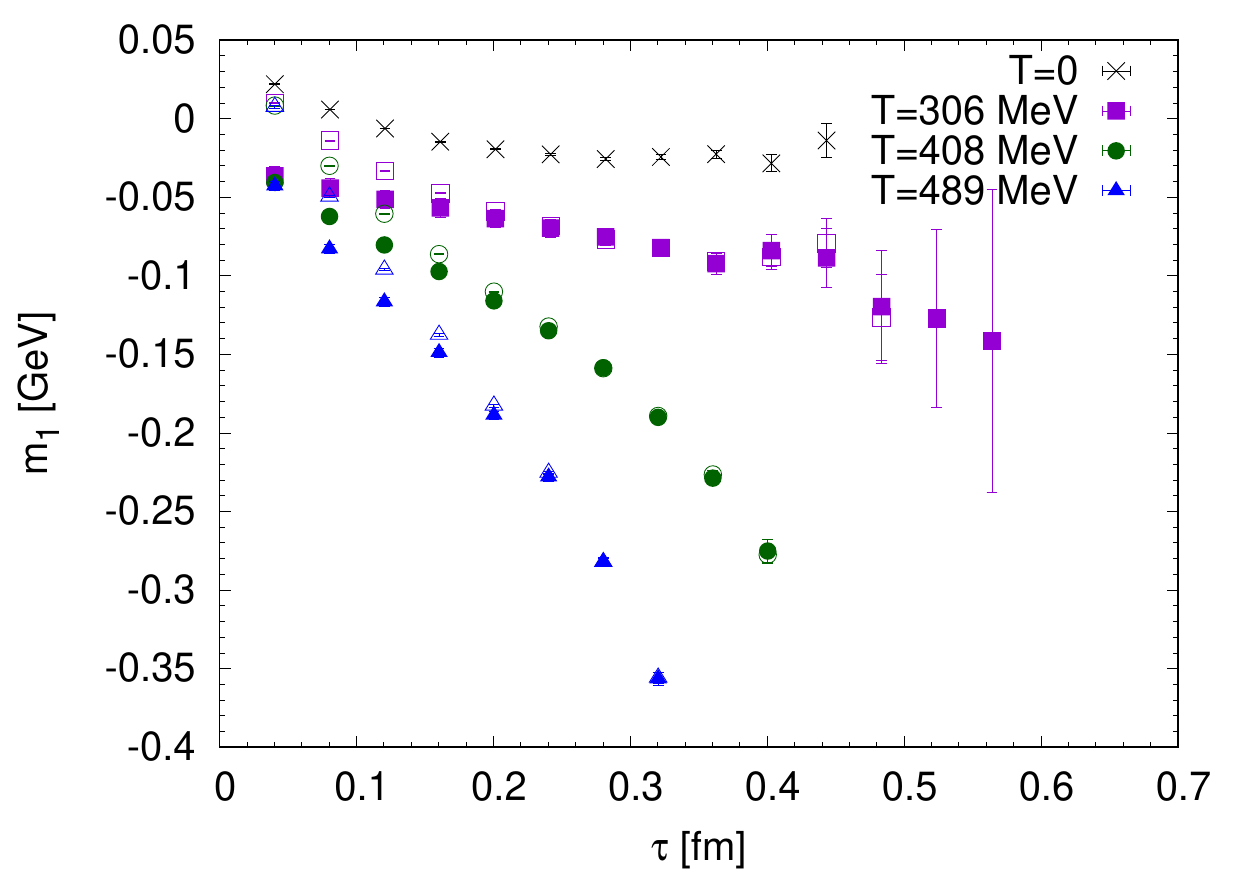}
\includegraphics[width=7cm]{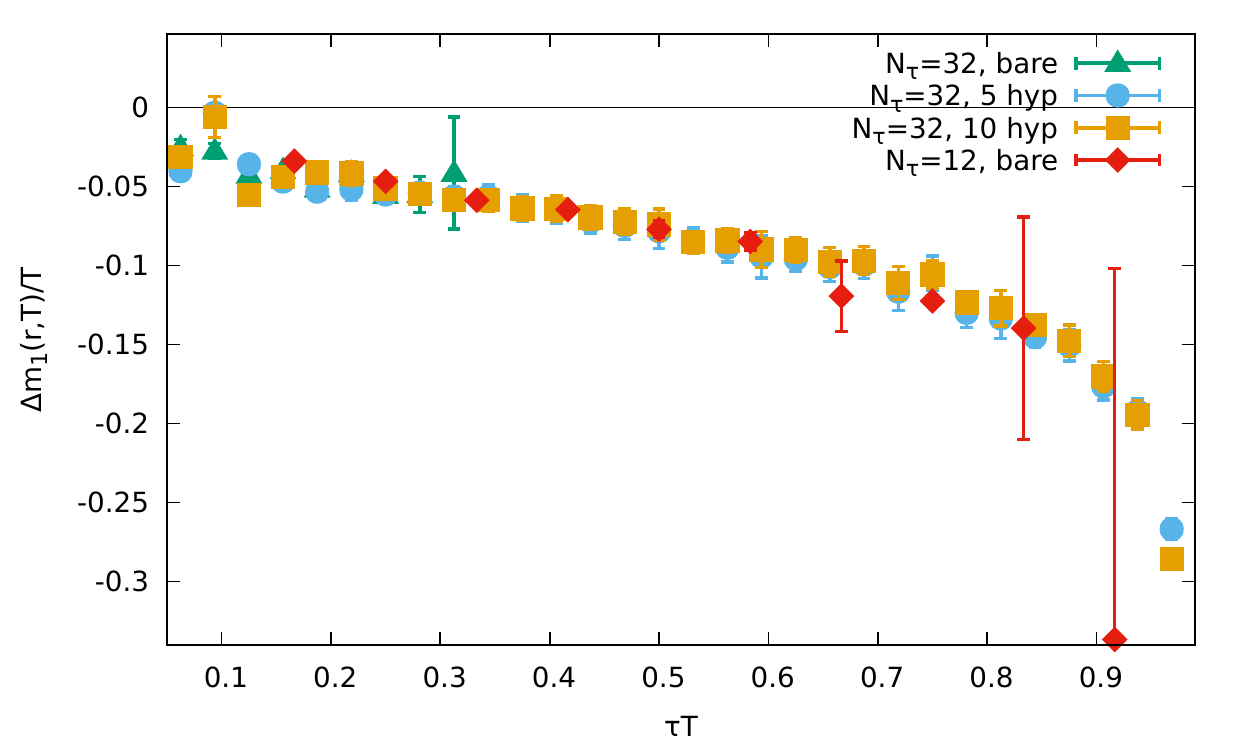}
\caption{Wilson line correlators in Coulomb gauge in (2+1)-flavor QCD. 
Correlators with smaller $N_\tau$ exhibit stronger $\tau$ dependence 
at large $\tau$, and lack the plateau present at $T=0$. 
Subtraction of the vacuum UV part (filled symbols) reveals an approximately 
linear decrease until $\tau \gtrsim \sfrac{\beta}{2}$ (left, 
from~\cite{Bala:2021fkm}). 
We see smearing independence as well as consistency with results 
obtained on coarser lattices at large enough $\tau T$ to be accessible 
by the latter. 
Smaller $\tau T$ values seem to suffer from major fluctuations due to 
large cancellations that become prominent after the subtraction of 
the large $T=0$ UV part (right). 
}
\label{fig:Wloop cumulants}
\end{figure}

In the following, we leave any finite heavy-quark mass behind and consider 
the static limit, which inherits from the NRQCD case the lack of symmetry 
under $\tau \to \beta-\tau$ and the simple Laplace kernel in the relation 
between correlator and spectral function in Eq.~\eqref{eq:inverse problem}. 
We define cumulants of the Wilson loop $W(r,T,\tau)$ (or, more generally, 
of any correlator $G(T,\tau)$) as 

\begin{align}
  m_1(r,T,\tau) &=
  -\frac{\partial \ln W(r,T,\tau)}{\partial \tau}, 
  \quad
  m_{n+1}(r,T,\tau) =\frac{\partial m_{n}(r,T,\tau)}{\partial \tau} \quad \text{for}~n>1.
  \label{eq:cumulants}
\end{align}

If $m_1$ is defined through a discrete derivative, e.g. the logarithm 
of a next-neighbor ratio, it coincides with the usual effective mass 
$m_1(r,T,\tau)= \sfrac{1}{a} \ln\left[ \sfrac{W(r,T,\tau)}{W(r,T,\tau+a)}\right]$. 
For the analysis of the Wilson loops -- which all represent extended 
operators -- the same technique of subtracting the vacuum UV part as 
in the NRQCD case has been applied, cf. Section~\ref{sec:NRQCD}. 
Results for $T=0$ or three different temperatures are shown in 
Fig.~\ref{fig:Wloop cumulants}, where the simplified $\tau$ dependence of 
the subtracted correlators at $\tau \lesssim \sfrac{\beta}{2}$ is evident. 
This study used the well-established HotQCD setup (HISQ/Tree action with 
$m_l=\sfrac{m_s}{20}$, see Section~\ref{sec:relativistic}), too. 

Instead of a rectangular Wilson loop, the authors considered the correlator 
of two temporal Wilson lines in Coulomb gauge, or a Wilson loop with spatially 
smeared (three-dimensional HYP smearing~\cite{Hasenfratz:2001hp}) spatial 
Wilson lines. 
It was demonstrated at $T=0$~\cite{Bazavov:2019qoo} that the smearing reduces 
the UV part of the correlator, and at $T>0$~\cite{Bala:2021fkm} that the 
smeared Wilson loops and Coulomb gauge Wilson line correlators are -- after 
appropriate subtraction of the respective vacuum UV parts -- statistically 
consistent in the range of approximately linear $\tau$ dependence, namely at 
$\tau \lesssim \sfrac{\beta}{2}$, but differ in their rapid falloff at 
$\tau \gtrsim \sfrac{\beta}{2}$ that is related to the low-energy tail. 
Hence, the behavior at $\tau \lesssim \sfrac{\beta}{2}$ may be 
understood as independent of the details of the interpolating operator.
This strongly points toward the existence of a quasiparticle state.

On the one hand, an approximately linear $\tau$ dependence of $m_1(r,T,\tau)$ 
for the UV-subtracted correlator implies nearly constant $m_2(r,T,\tau)$, 
and approximately zero $m_3(r,T,\tau)$ at $\tau \lesssim \sfrac{\beta}{2}$. 
On the other hand, the steep falloff at $\tau \gtrsim \sfrac{\beta}{2}$ implies 
that $m_2(r,T,\tau)$ becomes increasingly negative. 
This indicates negative $m_3(r,T,\tau)$ as well. 
So far, the correlators with $N_\tau \le 16$ did not provide sufficient 
resolution for more than the first three cumulants; $m_3(r,T,\tau)$ could be 
determined only for $T > 300\,\mr{MeV}$~\cite{Bala:2021fkm}. 
While most of the analysis focused on finite temperature lattices with 
$N_\tau=12$ close to the physical point ($m_l=\sfrac{m_s}{20}$, and a 
physical strange quark), the comparison to results with $N_\tau=16$ or~$10$ 
suggests mild cutoff effects. 
Juxtaposition with results for $m_l=\sfrac{m_s}{5}$ at 
$T \gtrsim 500\,\mr{MeV}$ suggests a mild quark-mass dependence (at high 
temperatures) as naively expected. 

In order to achieve a resolution of the higher cumulants capable of 
resolving an imprint of the non-trivial dynamics of the quasiparticle state, 
the HotQCD collaboration has commenced exploratory studies of Wilson line 
correlators in Coulomb gauge on exceptionally fine lattices 
$a \approx 0.028\,\mr{fm}$ using $N_\tau=32$ or $24$, i.e. in a fixed scale 
approach. 
In this study, the subtraction of the UV part -- previously implemented 
in terms of fits to the $T=0$ correlator -- was substituted by a direct 
calculation of the ratio between two correlators at different temperatures. 
The new approach yields in all studied cases statistically consistent 
results; the first cumulant of the ratio is the difference between first 
cumulants at the two different temperatures, i.e. 
$\Delta m_1(r,T) = m_1(r,T) - m_1(r,0)$, see Fig.~\ref{fig:Wloop cumulants}. 
Preliminary results have been reported at this 
conference~\cite{Hoying:2021mba}. 
Use of such fine lattices necessitated noise suppression techniques to obtain 
a good signal at intermediate to large $\tau$ even for quite small distances. 
The use of four-dimensional HYP smearing~\cite{Hasenfratz:2001hp} after 
Coulomb gauge fixing, which had been introduced for studying the screening 
masses of Polyakov loop correlators at large 
distances~\cite{Steinbeisser:2018sde, Petreczky:2021mef} 
is being used as a noise suppresion technique on these fine lattices as well. 
While the UV part cancels very well between smeared $T=0$ and finite temperature  
correlators, previously insignificant statistical fluctuations seem to 
become prominent due to the cancellations. 
For this reason, conclusive results on higher cumulants at 
$\tau \lesssim \sfrac{\beta}{2}$ may require improved statistics. 

A similar reasoning as in the non-relativstic case means that the fit in 
Eq.~\eqref{eq:gaussian+tail} could be applied to the UV subtracted Wilson 
loops as well. 
These fits have been reported at this conference~\cite{Parkar:2021wpv}, 
and suggest a strong-binding scenario, where the position of the lowest 
spectral feature $\Omega(r,T)$ is practically independent of the 
temperature, while the corresponding effective width $\Gamma(r,T)$ shows 
naive scaling $\sfrac{\Gamma(rT)}{T} = f(rT)$ for all temperatures down 
to the crossover. 
An alternative analysis following a different approach -- reported in 
Ref.~\cite{Parkar:2021wpv}, too -- also suggests a strong-binding scenario, 
where the position of the lowest spectral feature $\Omega(r,T)$ is 
nearly independent of the temperature, while the corresponding width 
$\Gamma(r,T)$ shows weaker than naive scaling with $T$ and $rT$. 
In this approach the Matsubara correlator at tree-level improved frequencies 
is obtained as the Fourier transform of the Euclidean-time correlator. 
Then the Matsubara correlator is interpolated using Pad\'e approximants, 
and finally analytically continued from imaginary to real frequencies, 
where the properties of the lowest spectral feature could be read off 
directly. 
The third approach discussed in Ref.~\cite{Parkar:2021wpv} is a Bayesian one, 
where the Bayesian Reconstruction (BR)~\cite{Burnier:2013nla} was applied 
as a regulator in the analysis of the Wilson loops. 
However, the very precise underlying lattice correlators on ensembles in 
the HotQCD setup are explicitly sensitive to tiny positivity violations at 
$r \sim a$ and $\tau \sim a$ due to the use of the improved gauge 
action; these can be resolved only on fine lattices~\cite{Bazavov:2019qoo}. 
Hence, the BR analysis was limited to sufficiently coarse lattices, namely 
for $N_\tau=12$ just at the crossover, where the BR method is consistent with the other results. 

Earlier Pad\'e or BR analyses on subsets of these data without some of the 
recent refinements of the analysis (for details, see 
Ref.~\cite{Bala:2021fkm}) found within large errors somewhat better support 
for the weak-binding picture~\cite{Petreczky:2018xuh}, similar to an 
earlier analysis using the quenched approximation or (2+1)-flavor lattice QCD 
with an AsqTad sea~\cite{Burnier:2014ssa}.

The leading-order HTL result in Eq.~\eqref{eq:Laine potential} could be 
a very suggestive inspiration for an analysis of the finite temperature Wilson loops 
on the lattice, since it contains a regularized Breit-Wigner peak. 
Such an approach had been followed in the past by fitting Wilson loops on 
the lattice with a rescaled form of this result~\cite{Bazavov:2014kva}. 
However, before one embarks on such an endeavor, it is cautious to discern
to which extent the lattice Wilson loops share characteristic features of 
the leading-order HTL result, as reported at this conference~\cite{Parkar:2021wpv}.  
The first cumulant of the correlator reconstructed from the HTL spectral 
function,

\begin{align}
  m_{1}^\mr{HTL}(r,\tau)
  &=\mr{Re}\,V_s^\mr{HTL}(r,T)+\frac{\mr{Im}\,V_s^\mr{HTL}(r,T)\beta}{\pi}\,
  \frac{\partial \ln \sin\left(\pi \frac{\tau}{\beta}\right)}
  {\partial \tau},
  \quad \text{for~} \tau \sim \frac{\beta}{2},
  \label{eq:HTL cumulant}
\end{align}

is antisymmetric under $\tau \to \beta-\tau$, and its midpoint value at 
$\tau=\sfrac{\beta}{2}$ is the real part $\mr{Re}\,V_s^\mr{HTL}(r,T)$, 
which coincides at the leading order with the singlet free energy in 
Coulomb gauge $F_S^\mr{HTL}(r,T)$. 
A divergence-free comparison is possible in terms of 
$\mr{Re}\,V_s(r,T)-F_S(r,T)$, since both share the same renormalon in the 
weak-coupling expansion or $\sfrac{1}{a}$ divergence in the lattice scheme. 
Due to the lack of a significant UV contribution in the HTL spectral function, 
however, juxtaposing the HTL expression with the lattice should be done 
in terms of the UV-subtracted data.  
First, the characteristic antisymmetry is not represented in the lattice 
data over a large range in $\tau$. 
Second, the decrease of the lattice data is much steeper than in HTL. 
Third, the midpoint value of the lattice result is only at very high 
temperatures $T \sim 2\,\mr{GeV}$ consistent with zero (the HTL prediction); 
see Refs.~\cite{Bala:2021fkm,Parkar:2021wpv} for further details. 

Nonetheless, with a fit range limited to a narrow interval around the 
midpoint $\tau=\sfrac{\beta}{2}$, it is legitimate to analyze the lattice 
Wilson loops using Eq.~\eqref{eq:HTL cumulant} in a discretized form 
and obtain predictions for $\Omega \sim \mr{Re}\,V_s$ and 
$\Gamma \sim \mr{Im}\,V_s$. 
In the past such an HTL-inspired analysis had been applied to 
spatially-smeared Wilson loops on anisotropic lattices in the 
quenched approximation~\cite{Bala:2019cqu}.
The results from such an analysis of the (2+1)-flavor QCD data in the HotQCD 
setup have been reported at this conference~\cite{Bala:2021jcw}. 
This analysis suggests a weak-binding scenario, where the position of the 
lowest spectral feature $\Omega(r,T)$ is quantitatively close to 
$F_S(r,T)$, while the corresponding width $\Gamma(r,T)$ shows 
weaker than naive scaling. 
In particular, $\Omega(r,T)$ stays close to $F_S(r,T)$ for $T>300\,\mr{MeV}$ 
much beyond $rT \lesssim 0.3$ where similar agreement has been 
observed for the $T=0$ static energy as well~\cite{Bazavov:2018wmo}. 
These findings are similar to the earlier ones in the quenched approximation~\cite{Bala:2019cqu}. 

All recently applied lattice analysis methods suggest that the finite 
temperature Wilson loop exhibits a well-defined lowest spectral feature 
$\Omega(r,T)$ at $r \lesssim \sfrac{1.5}{T}$ and up to $T \sim 2\,\mr{GeV}$, 
which can be considered as a complex, quasiparticle remnant of 
the $T=0$ static energy. 
All the methods robustly point towards a significant width of this 
spectral feature -- or in other words to a large imaginary part 
of the complex static energy. 
This width grows for increasing $r$ or increasing temperature, 
but can be seen already at the crossover. 
However, at present static quark-antiquark correlators do not provide a 
conclusive picture on the question of weak- vs strong-binding scenario yet. 
The expected screening of the real part of the static energy is not 
evident in our calculations. 

\subsection{Heavy quarkonia free energies}
\label{sec:free energies}

The free energies of heavy quarkonia can be studied by evaluating the 
thermal static quark-antiquark correlators at $\tau=\beta$, i.e., in 
a gauge-invariant manner in terms of the Polyakov loop correlator

\begin{align}
  C_P(r,T) = \braket{ P(0) P^\dagger(\bm{r}) }, 
  \label{eq:PLC}
\end{align}

with $P(\bm{r})=\braket{\mathrm{Re\,Tr} U(\beta,\bm{r})}$ being the trace of the 
Polyakov loop. 
After renormalization $C_P(r,T) = e^{-\sfrac{F_{q\bar q}(r,T)}{T}}$. 
For $r \ll \beta$ this correlator can be expressed in terms of the $T=0$ 
singlet- or octet-potentials and the adjoint Polyakov loop; 
substituting the static energy on the lattice instead of the potentials 
and accounting for Casimir scaling violations, 
good agreement up $rT \lesssim 0.35$ suggests that within such distances 
no thermal modification occurs besides the changes of the adjoint 
Polyakov loop~\cite{Bazavov:2018wmo}. 
The increasing value of the adjoint Polyakov loop implies that color-octet 
states -- dissociated static quarkonia -- become less suppressed. 

At large distances, $C_P(r,T) = C_P^R(r,T) + C_P^I(r,T)$ can be understood 
as the sum of contributions from the real or imaginary parts of the Polyakov 
loops, while it asymptotes at $\braket{P}^2$ in the large distance limit. 
Their ratio

\begin{align}
  C_P^R(r,T) = \braket{ \mathrm{Re\,}P(0) \mathrm{Re\,}P^\dagger(\bm{r}) }, 
  \quad
  C_P^I(r,T) = \braket{ \mathrm{Im\,}P(0) \mathrm{Im\,}P^\dagger(\bm{r}) }, 
  \label{eq:PLCRI}
\end{align}

where $C_P^R(r,T)$ is dominated by two-gluon exchange, while $C_P^I(r,T)$ is 
dominated by three-gluon exchange. 
Asymptotically, the normalized correlators 

\begin{align}
  \widetilde{C}_P^R(r,T) = \frac{\braket{ \mathrm{Re\,}P(0) \mathrm{Re\,}P^\dagger(\bm{r}) } - \braket{P}^2}{\braket{P}^2}, 
  \quad
  \widetilde{C}_P^I(r,T) = \frac{\braket{ \mathrm{Im\,}P(0) \mathrm{Im\,}P^\dagger(\bm{r}) }}{\braket{P}^2}, 
  \label{eq:PLCRIN}
\end{align}

decay as $\widetilde{C}_P^{R,I}(r,T) \propto \sfrac{e^{m_{R,I}r}}{rT}$ with 
screening masses $m_{R,I}$. 
At leading order, these screening masses satisfy $m_R=2\md$, $m_I=3\md$; above 
the crossover region temperature they are proportional to the temperature up 
to subleading, logarithmic corrections. 

Computing the Polyakov loop correlators at large distances is challenging 
due to the poor signal-to-noise ratio. 
However, using four-dimensional hypercubic smearing~\cite{Hasenfratz:2001hp} 
a good signal can be obtained up to $r \sim \beta$ in the HotQCD setup (see Sec.~\ref{sec:relativistic}), which is sufficient to determine the 
screening masses $\sfrac{m_R}{T} \simeq 4.5$ and $\sfrac{m_I}{T} \simeq 8$ 
over the temperature range $(170-1500)\,\mathrm{MeV}$ that were reported 
at this  conference~\cite{Petreczky:2021mef}. 
These results are quantitatively consistent with earlier calculations in 
full QCD~\cite{Borsanyi:2015yka} or in dimensionally reduced effective 
field theory~\cite{Hart:2000ha}. 
The corresponding screening length 
$\sfrac{1}{\md} \simeq \sfrac{2}{m_R} \simeq \sfrac{3}{m_I}$ is 
$(0.38-0.44)/T$, which would suggest melting temperatures due to screening 
of about $380\,\mathrm{MeV}$ or $190\,\mathrm{MeV}$ for 
$\mathrm{\Upsilon(1S)}$ or $\mathrm{J/\Psi}$ , respectively.

\section{Conclusions}
\label{sec:conclusions}

In these proceedings, we have reviewed the recent advances in lattice 
gauge theory with respect to the non-pertubative calculation of 
in-medium heavy-quark observables at finite temperature. 
We have inspected two different, but interrelated problems -- isolated 
heavy quarks that are subject to dynamical \emph{transport phenomena}, 
and quarkonia, i.e., correlated heavy quark-antiquark pairs, that are 
subject to dynamical \emph{melting}. 
The dynamical nature of both phenomena implies that the ill-posed 
inverse problem in Eq.~\eqref{eq:inverse problem} needs to be solved. 

The key observables for heavy-quark transport are chromo-electric and 
-magnetic field-strength correlators, which require efficient noise 
suppression techniques and have been limited so far to the quenched 
approximation.  
Recently, they have been determined using the gradient flow instead 
of the multi-level algorithm for the noise reduction as reported at 
this conference~\cite{Altenkort:2021ntw, Mayer-Steudte:2021hei}. 
This alleviates issues with renormalization in previous 
calculations, and opens up the prospect of obtaining the first 
non-perturbative QCD results in the near future. 

The key observables for in-medium heavy quarkonia are either spatial or 
temporal quarkonia correlators, or thermal Wilson loops, which have been 
studied with three dynamical flavors by multiple groups. 
Spatial bottomonia correlators have been computed~\cite{Petreczky:2021zmz} 
with HISQ action in an extension of a previous study in the charm 
sector~\cite{Bazavov:2014cta}. 
Both studies support the expected melting temperatures of a 
strong-binding scenario as reported at this conference~\cite{Petreczky:2021hyd}.  
Preliminary results from a study of spatial correlators in the chiral limit 
were reported at this conference~\cite{lat2021laudicina}, which determined 
that the next-to-leading order does not quantitatively capture the 
percent-level deviation between the leading order and the lattice even at 
the electroweak scale. 
Of course, these findings apply to heavy quarkonia as well. 
Temporal bottomonia correlators have been computed using NRQCD by the 
FASTSUM and HotQCD collaborations. 
The HotQCD results~\cite{Larsen:2019bwy, Larsen:2019zqv, Larsen:2020rjk, 
Shi:2021qri} using extended sources generally support the strong-binding 
scenario. 
Older results from FASTSUM~\cite{Aarts:2014cda} or HotQCD~\cite{Kim:2018yhk} 
as well as preliminary FASTSUM results~\cite{Spriggs:2021jsh, 
Spriggs:2021ieo, Offler:2021fmg, Page:2021ohe, Spriggs:2021dsb} 
with local point sources paint a less conclusive picture. 
Thermal Wilson loops or Coulomb gauge Wilson line correlators have been 
analyzed by the HotQCD collaboration~\cite{Bala:2021fkm}. 
These calculations robustly point to the existence of a well-separated,  
lowest spectral structure with a finite width as the remnant of the $T=0$ 
static energy, but are inconclusive with regard to the 
weak- or strong-binding scenarios. 
Some calculations using the same data were reported at this conference, which 
supported for an HTL-inspired Ansatz the weak-binding~\cite{Bala:2021jcw}, or 
for model fits or Pad\'e analysis the strong-binding scenario~\cite{Parkar:2021wpv}. 
The screening length obtained from Polyakov correlators would suggest melting 
temperatures closer to a strong-binding scenario~\cite{Petreczky:2021mef}. 
In this regard it is not obvious how this result, together with the robust 
evidence for the finite thermal width, and the melting patterns from the 
spatial quarkonia correlators can be reconciled in a consistent picture. 
The most likely path towards a more conclusive result is the use of very 
fine lattices that could exclude some solutions of the inverse problem; 
preliminary results in this direction from the HotQCD collaboration have been 
reported at this conference~\cite{Hoying:2021mba}. 
Of course, the required fine ensembles would be valuable for calculation of 
spatial or temporal quarkonia correlators, or for the field-strength 
correlators from which transport coefficients are determined. 

\section*{Acknowledgments}

We thank P.~Petreczky for discussions, careful reading and comments 
on the manuscript. 
J.H.W.’s research was funded by the Deutsche Forschungsgemeinschaft 
(DFG, German Research Foundation) - Projektnummer 417533893/GRK2575 
``Rethinking Quantum Field Theory''.


\begin{thebibliography}{99}

%%%%%%%%%%%%%%%%%%%%%%%%%%%%%%%%%%%%%%%%%%%%%%%%%%%%%%%%%%%
%% Introduction
%%%%%%%%%%%%%%%%%%%%%%%%%%%%%%%%%%%%%%%%%%%%%%%%%%%%%%%%%%%

\bibitem{Akiba:2015jwa}
Y.~Akiba, A.~Angerami, H.~Caines, A.~Frawley, U.~Heinz, B.~Jacak, J.~Jia, T.~Lappi, W.~Li and A.~Majumder, \textit{et al.}
%``The Hot QCD White Paper: Exploring the Phases of QCD at RHIC and the LHC,''
[arXiv:1502.02730 [nucl-ex]].

\bibitem{Ding:2016qdj}
H.~T.~Ding, F.~Karsch and S.~Mukherjee,
%``Thermodynamics of Strong-Interaction Matter from Lattice QCD,''
doi:10.1142/9789814663717\_0001.

\bibitem{Bazavov:2019lgz}
A.~Bazavov \textit{et al.} [USQCD],
%``Hot-dense Lattice QCD: USQCD whitepaper 2018,''
Eur. Phys. J. A \textbf{55}, no.11, 194 (2019)
% doi:10.1140/epja/i2019-12922-0
[arXiv:1904.09951 [hep-lat]].

\bibitem{Bazavov:2020teh}
A.~Bazavov and J.~H.~Weber,
%``Color Screening in Quantum Chromodynamics,''
Prog. Part. Nucl. Phys. \textbf{116}, 103823 (2021)
% doi:10.1016/j.ppnp.2020.103823
[arXiv:2010.01873 [hep-lat]].

\bibitem{Gross:1980br}
D.~J.~Gross, R.~D.~Pisarski and L.~G.~Yaffe,
%``QCD and Instantons at Finite Temperature,''
Rev. Mod. Phys. \textbf{53}, 43 (1981).
% doi:10.1103/RevModPhys.53.43

\bibitem{Karsch:2003vd}
F.~Karsch, K.~Redlich and A.~Tawfik,
%``Hadron resonance mass spectrum and lattice QCD thermodynamics,''
Eur. Phys. J. C \textbf{29}, 549-556 (2003)
% doi:10.1140/epjc/s2003-01228-y
[arXiv:hep-ph/0303108 [hep-ph]].

\bibitem{Hagedorn:1965st}
R.~Hagedorn,
%``Statistical thermodynamics of strong interactions at high-energies,''
Nuovo Cim. Suppl. \textbf{3}, 147-186 (1965)
CERN-TH-520.

%%%%%%%%%%%%%%%%%%%%%%%%%%%%%%%%%%%%%%%%%%%%%%%%%%%%%%%%%%%

\bibitem{Cabibbo:1975ig}
N.~Cabibbo and G.~Parisi,
%``Exponential Hadronic Spectrum and Quark Liberation,''
Phys. Lett. B \textbf{59}, 67-69 (1975).
% doi:10.1016/0370-2693(75)90158-6

\bibitem{HotQCD:2019xnw}
H.~T.~Ding \textit{et al.} [HotQCD],
%``Chiral Phase Transition Temperature in ( 2+1 )-Flavor QCD,''
Phys. Rev. Lett. \textbf{123}, no.6, 062002 (2019)
% doi:10.1103/PhysRevLett.123.062002
[arXiv:1903.04801 [hep-lat]].

\bibitem{HotQCD:2018pds}
A.~Bazavov \textit{et al.} [HotQCD],
%``Chiral crossover in QCD at zero and non-zero chemical potentials,''
Phys. Lett. B \textbf{795}, 15-21 (2019)
% doi:10.1016/j.physletb.2019.05.013
[arXiv:1812.08235 [hep-lat]].

\bibitem{Borsanyi:2020fev}
S.~Borsanyi, Z.~Fodor, J.~N.~Guenther, R.~Kara, S.~D.~Katz, P.~Parotto, A.~Pasztor, C.~Ratti and K.~K.~Szabo,
%``QCD Crossover at Finite Chemical Potential from Lattice Simulations,''
Phys. Rev. Lett. \textbf{125}, no.5, 052001 (2020)
[arXiv:2002.02821 [hep-lat]].

\bibitem{Bazavov:2018wmo}
A.~Bazavov \textit{et al.} [TUMQCD],
%``Color screening in (2+1)-flavor QCD,''
Phys. Rev. D \textbf{98}, no.5, 054511 (2018)
% doi:10.1103/PhysRevD.98.054511
[arXiv:1804.10600 [hep-lat]].

%%%%%%%%%%%%%%%%%%%%%%%%%%%%%%%%%%%%%%%%%%%%%%%%%%%%%%%%%%%

\bibitem{Brambilla:2010cs}
N.~Brambilla, S.~Eidelman, B.~K.~Heltsley, R.~Vogt, G.~T.~Bodwin, E.~Eichten, A.~D.~Frawley, A.~B.~Meyer, R.~E.~Mitchell and V.~Papadimitriou, \textit{et al.}
%``Heavy Quarkonium: Progress, Puzzles, and Opportunities,''
Eur. Phys. J. C \textbf{71}, 1534 (2011)
% doi:10.1140/epjc/s10052-010-1534-9
[arXiv:1010.5827 [hep-ph]].

\bibitem{Rothkopf:2019ipj}
A.~Rothkopf,
%``Heavy Quarkonium in Extreme Conditions,''
Phys. Rept. \textbf{858}, 1-117 (2020)
% doi:10.1016/j.physrep.2020.02.006
[arXiv:1912.02253 [hep-ph]].

\bibitem{Brambilla:2004jw}
N.~Brambilla, A.~Pineda, J.~Soto and A.~Vairo,
%``Effective Field Theories for Heavy Quarkonium,''
Rev. Mod. Phys. \textbf{77}, 1423 (2005)
% doi:10.1103/RevModPhys.77.1423
[arXiv:hep-ph/0410047 [hep-ph]].

%%%%%%%%%%%%%%%%%%%%%%%%%%%%%%%%%%%%%%%%%%%%%%%%%%%%%%%%%%%

\bibitem{Matsui:1986dk}
T.~Matsui and H.~Satz,
%``$J/\psi$ Suppression by Quark-Gluon Plasma Formation,''
Phys. Lett. B \textbf{178}, 416-422 (1986).
% doi:10.1016/0370-2693(86)91404-8

\bibitem{Karsch:2005nk}
F.~Karsch, D.~Kharzeev and H.~Satz,
%``Sequential charmonium dissociation,''
Phys. Lett. B \textbf{637}, 75-80 (2006)
% doi:10.1016/j.physletb.2006.03.078
[arXiv:hep-ph/0512239 [hep-ph]].

%%%%%%%%%%%%%%%%%%%%%%%%%%%%%%%%%%%%%%%%%%%%%%%%%%%%%%%%%%%

\bibitem{Svetitsky:1987gq}
B.~Svetitsky,
%``Diffusion of charmed quarks in the quark-gluon plasma,''
Phys. Rev. D \textbf{37}, 2484-2491 (1988).
% doi:10.1103/PhysRevD.37.2484.

\bibitem{Meyer:2010tt}
H.~B.~Meyer,
%``The errant life of a heavy quark in the quark-gluon plasma,''
New J. Phys. \textbf{13}, 035008 (2011)
% doi:10.1088/1367-2630/13/3/035008
[arXiv:1012.0234 [hep-lat]].

\bibitem{Banerjee:2011ra}
D.~Banerjee, S.~Datta, R.~Gavai and P.~Majumdar,
%``Heavy Quark Momentum Diffusion Coefficient from Lattice QCD,''
Phys. Rev. D \textbf{85}, 014510 (2012)
% doi:10.1103/PhysRevD.85.014510
[arXiv:1109.5738 [hep-lat]].

\bibitem{Francis:2015daa}
A.~Francis, O.~Kaczmarek, M.~Laine, T.~Neuhaus and H.~Ohno,
%``Nonperturbative estimate of the heavy quark momentum diffusion coefficient,''
Phys. Rev. D \textbf{92}, no.11, 116003 (2015)
% doi:10.1103/PhysRevD.92.116003
[arXiv:1508.04543 [hep-lat]].

\bibitem{Brambilla:2020siz}
N.~Brambilla, V.~Leino, P.~Petreczky and A.~Vairo,
%``Lattice QCD constraints on the heavy quark diffusion coefficient,''
Phys. Rev. D \textbf{102}, no.7, 074503 (2020)
% doi:10.1103/PhysRevD.102.074503
[arXiv:2007.10078 [hep-lat]].

\bibitem{Rapp:2009my}
R.~Rapp and H.~van Hees,
%``Heavy Quarks in the Quark-Gluon Plasma,''
% doi:10.1142/9789814293297\_0003
[arXiv:0903.1096 [hep-ph]].

\bibitem{Laine:2011is}
M.~Laine,
%``Heavy flavour kinetic equilibration in the confined phase,''
JHEP \textbf{04}, 124 (2011)
% doi:10.1007/JHEP04(2011)124
[arXiv:1103.0372 [hep-ph]].

\bibitem{Herzog:2006gh}
C.~P.~Herzog, A.~Karch, P.~Kovtun, C.~Kozcaz and L.~G.~Yaffe,
%``Energy loss of a heavy quark moving through N=4 supersymmetric Yang-Mills plasma,''
JHEP \textbf{07}, 013 (2006)
% doi:10.1088/1126-6708/2006/07/013
[arXiv:hep-th/0605158 [hep-th]].

%%%%%%%%%%%%%%%%%%%%%%%%%%%%%%%%%%%%%%%%%%%%%%%%%%%%%%%%%%%
%% Heavy-quark transport
%%%%%%%%%%%%%%%%%%%%%%%%%%%%%%%%%%%%%%%%%%%%%%%%%%%%%%%%%%%

\bibitem{Bouttefeux:2020ycy}
A.~Bouttefeux and M.~Laine,
%``Mass-suppressed effects in heavy quark diffusion,''
JHEP \textbf{12}, 150 (2020)
doi:10.1007/JHEP12(2020)150
[arXiv:2010.07316 [hep-ph]].
%7 citations counted in INSPIRE as of 27 Nov 2021

\bibitem{Caron-Huot:2009ncn}
S.~Caron-Huot, M.~Laine and G.~D.~Moore,
%``A Way to estimate the heavy quark thermalization rate from the lattice,''
JHEP \textbf{04}, 053 (2009)
% doi:10.1088/1126-6708/2009/04/053
[arXiv:0901.1195 [hep-lat]].

\bibitem{Christensen:2016wdo}
C.~Christensen and M.~Laine,
%``Perturbative renormalization of the electric field correlator,''
Phys. Lett. B \textbf{755}, 316-323 (2016)
% doi:10.1016/j.physletb.2016.02.020
[arXiv:1601.01573 [hep-lat]].

%%%%%%%%%%%%%%%%%%%%%%%%%%%%%%%%%%%%%%%%%%%%%%%%%%%%%%%%%%%

\bibitem{Luscher:2010iy}
M.~L\"uscher,
%``Properties and uses of the Wilson flow in lattice QCD,''
JHEP \textbf{08}, 071 (2010)
[erratum: JHEP \textbf{03}, 092 (2014)]
% doi:10.1007/JHEP08(2010)071
[arXiv:1006.4518 [hep-lat]].

%%%%%%%%%%%%%%%%%%%%%%%%%%%%%%%%%%%%%%%%%%%%%%%%%%%%%%%%%%%

\bibitem{Altenkort:2020fgs}
L.~Altenkort, A.~M.~Eller, O.~Kaczmarek, L.~Mazur, G.~D.~Moore and H.~T.~Shu,
%``Heavy quark momentum diffusion from the lattice using gradient flow,''
Phys. Rev. D \textbf{103}, no.1, 014511 (2021)
% doi:10.1103/PhysRevD.103.014511
[arXiv:2009.13553 [hep-lat]].

\bibitem{Mayer-Steudte:2021hei}
J.~Mayer-Steudte, N.~Brambilla, V.~Leino and P.~Petreczky,
%``Chromoelectric and chromomagnetic correlators at high temperature from gradient flow,''
[arXiv:2111.10340 [hep-lat]].

\bibitem{Altenkort:2021ntw}
L.~Altenkort, A.~M.~Eller, O.~Kaczmarek, L.~Mazur, G.~D.~Moore and H.~T.~Shu,
%``Continuum extrapolation of the gradient-flowed color-magnetic correlator at $1.5\,T_c$,''
[arXiv:2111.12462 [hep-lat]].

\bibitem{Brambilla:2019tpt}
N.~Brambilla, M.~A.~Escobedo, A.~Vairo and P.~Vander Griend,
%``Transport coefficients from in medium quarkonium dynamics,''
Phys. Rev. D \textbf{100}, no.5, 054025 (2019)
% doi:10.1103/PhysRevD.100.054025
[arXiv:1903.08063 [hep-ph]].

%%%%%%%%%%%%%%%%%%%%%%%%%%%%%%%%%%%%%%%%%%%%%%%%%%%%%%%%%%%
%% Heavy quarkonia
%%%%%%%%%%%%%%%%%%%%%%%%%%%%%%%%%%%%%%%%%%%%%%%%%%%%%%%%%%%

\bibitem{Manohar:1983md}
A.~Manohar and H.~Georgi,
%``Chiral Quarks and the Nonrelativistic Quark Model,''
Nucl. Phys. B \textbf{234}, 189-212 (1984).
% doi:10.1016/0550-3213(84)90231-1

%%%%%%%%%%%%%%%%%%%%%%%%%%%%%%%%%%%%%%%%%%%%%%%%%%%%%%%%%%%

\bibitem{Brambilla:1999xf}
N.~Brambilla, A.~Pineda, J.~Soto and A.~Vairo,
%``Potential NRQCD: An Effective theory for heavy quarkonium,''
Nucl. Phys. B \textbf{566}, 275 (2000)
% doi:10.1016/S0550-3213(99)00693-8
[arXiv:hep-ph/9907240 [hep-ph]].

%%%%%%%%%%%%%%%%%%%%%%%%%%%%%%%%%%%%%%%%%%%%%%%%%%%%%%%%%%%

\bibitem{Burnier:2012az}
Y.~Burnier and A.~Rothkopf,
%``Disentangling the timescales behind the non-perturbative heavy quark potential,''
Phys. Rev. D \textbf{86}, 051503 (2012)
% doi:10.1103/PhysRevD.86.051503
[arXiv:1208.1899 [hep-ph]].

%%%%%%%%%%%%%%%%%%%%%%%%%%%%%%%%%%%%%%%%%%%%%%%%%%%%%%%%%%%

\bibitem{Laine:2006ns}
M.~Laine, O.~Philipsen, P.~Romatschke and M.~Tassler,
%``Real-time static potential in hot QCD,''
JHEP \textbf{03}, 054 (2007)
% doi:10.1088/1126-6708/2007/03/054
[arXiv:hep-ph/0611300 [hep-ph]].

\bibitem{Brambilla:2008cx}
N.~Brambilla, J.~Ghiglieri, A.~Vairo and P.~Petreczky,
%``Static quark-antiquark pairs at finite temperature,''
Phys. Rev. D \textbf{78}, 014017 (2008)
% doi:10.1103/PhysRevD.78.014017
[arXiv:0804.0993 [hep-ph]].

%%%%%%%%%%%%%%%%%%%%%%%%%%%%%%%%%%%%%%%%%%%%%%%%%%%%%%%%%%%
%% Screening correlators
%%%%%%%%%%%%%%%%%%%%%%%%%%%%%%%%%%%%%%%%%%%%%%%%%%%%%%%%%%%

\bibitem{Karsch:2012na}
F.~Karsch, E.~Laermann, S.~Mukherjee and P.~Petreczky,
%``Signatures of charmonium modification in spatial correlation functions,''
Phys. Rev. D \textbf{85}, 114501 (2012)
% doi:10.1103/PhysRevD.85.114501
[arXiv:1203.3770 [hep-lat]].

\bibitem{Bazavov:2014cta}
A.~Bazavov, F.~Karsch, Y.~Maezawa, S.~Mukherjee and P.~Petreczky,
%``In-medium modifications of open and hidden strange-charm mesons from spatial correlation functions,''
Phys. Rev. D \textbf{91}, no.5, 054503 (2015)
% doi:10.1103/PhysRevD.91.054503
[arXiv:1411.3018 [hep-lat]].

\bibitem{Bazavov:2019www}
A.~Bazavov, S.~Dentinger, H.~T.~Ding, P.~Hegde, O.~Kaczmarek, F.~Karsch, E.~Laermann, A.~Lahiri, S.~Mukherjee and H.~Ohno, \textit{et al.}
%``Meson screening masses in (2+1)-flavor QCD,''
Phys. Rev. D \textbf{100}, no.9, 094510 (2019)
% doi:10.1103/PhysRevD.100.094510
[arXiv:1908.09552 [hep-lat]].

\bibitem{Petreczky:2021zmz}
P.~Petreczky, S.~Sharma and J.~H.~Weber,
%``Bottomonium melting from screening correlators at high temperature,''
Phys. Rev. D \textbf{104}, no.5, 054511 (2021)
% doi:10.1103/PhysRevD.104.054511
[arXiv:2107.11368 [hep-lat]].

%\cite{Weber:2021hro}
\bibitem{Weber:2021hro}
J.~H.~Weber, A.~Bazavov and P.~Petreczky,
%``Update on (2+1+1)-flavor QCD equation of state,''
PoS \textbf{LATTICE2021}, 060 (2021)
[arXiv:2110.03606 [hep-lat]].
%0 citations counted in INSPIRE as of 29 Nov 2021

\bibitem{Petreczky:2021hyd}
P.~Petreczky, S.~Sharma and J.~H.~Weber,
%``Bottomonia screening masses from $2 + 1$ flavor QCD,''
[arXiv:2112.07043 [hep-lat]].

\bibitem{lat2021laudicina}
D.~Laudicina, et al.,
% `` Computation of QCD meson screening masses at high temperature ,''
PoS(LATTICE2021) 190.

%%%%%%%%%%%%%%%%%%%%%%%%%%%%%%%%%%%%%%%%%%%%%%%%%%%%%%%%%%%
%% NRQCD correlators
%%%%%%%%%%%%%%%%%%%%%%%%%%%%%%%%%%%%%%%%%%%%%%%%%%%%%%%%%%%

\bibitem{Thacker:1990bm}
B.~A.~Thacker and G.~P.~Lepage,
%``Heavy quark bound states in lattice QCD,''
Phys. Rev. D \textbf{43}, 196-208 (1991).
% doi:10.1103/PhysRevD.43.196

\bibitem{Lepage:1992tx}
G.~P.~Lepage, L.~Magnea, C.~Nakhleh, U.~Magnea and K.~Hornbostel,
%``Improved nonrelativistic QCD for heavy quark physics,''
Phys. Rev. D \textbf{46}, 4052-4067 (1992)
% doi:10.1103/PhysRevD.46.4052
[arXiv:hep-lat/9205007 [hep-lat]].

%%%%%%%%%%%%%%%%%%%%%%%%%%%%%%%%%%%%%%%%%%%%%%%%%%%%%%%%%%%

\bibitem{Aarts:2014cda}
G.~Aarts, C.~Allton, T.~Harris, S.~Kim, M.~P.~Lombardo, S.~M.~Ryan and J.~I.~Skullerud,
%``The bottomonium spectrum at finite temperature from N$_{f}$ = 2 + 1 lattice QCD,''
JHEP \textbf{07}, 097 (2014)
% doi:10.1007/JHEP07(2014)097
[arXiv:1402.6210 [hep-lat]].

\bibitem{Kim:2018yhk}
S.~Kim, P.~Petreczky and A.~Rothkopf,
%``Quarkonium in-medium properties from realistic lattice NRQCD,''
JHEP \textbf{11}, 088 (2018)
% doi:10.1007/JHEP11(2018)088
[arXiv:1808.08781 [hep-lat]].

\bibitem{Larsen:2019bwy}
R.~Larsen, S.~Meinel, S.~Mukherjee and P.~Petreczky,
%``Thermal broadening of bottomonia: Lattice nonrelativistic QCD with extended operators,''
Phys. Rev. D \textbf{100}, no.7, 074506 (2019)
% doi:10.1103/PhysRevD.100.074506
[arXiv:1908.08437 [hep-lat]].

\bibitem{Larsen:2019zqv}
R.~Larsen, S.~Meinel, S.~Mukherjee and P.~Petreczky,
%``Excited bottomonia in quark-gluon plasma from lattice QCD,''
Phys. Lett. B \textbf{800}, 135119 (2020)
% doi:10.1016/j.physletb.2019.135119
[arXiv:1910.07374 [hep-lat]].

\bibitem{Larsen:2020rjk}
R.~Larsen, S.~Meinel, S.~Mukherjee and P.~Petreczky,
%``Bethe-Salpeter amplitudes of Upsilons,''
Phys. Rev. D \textbf{102}, 114508 (2020)
% doi:10.1103/PhysRevD.102.114508
[arXiv:2008.00100 [hep-lat]].

%%%%%%%%%%%%%%%%%%%%%%%%%%%%%%%%%%%%%%%%%%%%%%%%%%%%%%%%%%%

\bibitem{Spriggs:2021jsh}
T.~Spriggs, G.~Aarts, C.~Allton, T.~Burns, B.~J\"ager, S.~Kim, M.~P.~Lombardo, S.~Offler, B.~Page and S.~M.~Ryan, \textit{et al.}
%``Bottomonium spectral widths at nonzero temperature using maximum likelihood,''
[arXiv:2112.01599 [hep-lat]].
%0 citations counted in INSPIRE as of 07 Dec 2021

\bibitem{Page:2021ohe}
B.~Page, G.~Aarts, C.~Allton, B.~J\"ager, S.~Kim, M.~P.~Lombardo, S.~Offler, S.~M.~Ryan, J.~I.~Skullerud and T.~Spriggs,
%``Spectral reconstruction in NRQCD via the Backus-Gilbert method,''
[arXiv:2112.02075 [hep-lat]].

\bibitem{Offler:2021fmg}
S.~Offler, G.~Aarts, C.~Allton, B.~J\"ager, S.~Kim, M.~P.~Lombardo, B.~Page, S.~M.~Ryan, J.~I.~Skullerud and T.~Spriggs,
%``Reconstruction of bottomonium spectral functions in thermal QCD using Kernel Ridge Regression,''
[arXiv:2112.02116 [hep-lat]].

\bibitem{Spriggs:2021dsb}
T.~Spriggs, G.~Aarts, C.~Allton, T.~Burns, R.~H.~D'Arcy, B.~J\"ager, S.~Kim, M.~P.~Lombardo, S.~Offler and B.~Page, \textit{et al.}
%``A comparison of spectral reconstruction methods applied to non-zero temperature NRQCD meson correlation functions,''
[arXiv:2112.04201 [hep-lat]].

%%%%%%%%%%%%%%%%%%%%%%%%%%%%%%%%%%%%%%%%%%%%%%%%%%%%%%%%%%%

\bibitem{Petreczky:2017aiz}
P.~Petreczky \textit{et al.} [TUMQCD],
%``Lattice Calculations of Heavy Quark Potential at Finite Temperature,''
Nucl. Phys. A \textbf{967}, 592-595 (2017)
% doi:10.1016/j.nuclphysa.2017.04.011
[arXiv:1704.08573 [hep-lat]].

\bibitem{Bala:2021fkm}
D.~Bala, O.~Kaczmarek, R.~Larsen, S.~Mukherjee, G.~Parkar, P.~Petreczky, A.~Rothkopf and J.~H.~Weber,
%``Static quark anti-quark interactions at non-zero temperature from lattice QCD,''
[arXiv:2110.11659 [hep-lat]].

%%%%%%%%%%%%%%%%%%%%%%%%%%%%%%%%%%%%%%%%%%%%%%%%%%%%%%%%%%%

\bibitem{Shi:2021qri}
S.~Shi, K.~Zhou, J.~Zhao, S.~Mukherjee and P.~Zhuang,
%``Heavy Quark Potential in QGP: DNN meets LQCD,''
[arXiv:2105.07862 [hep-ph]].

\bibitem{Spriggs:2021ieo}
T.~Spriggs, C.~Allton, T.~Burns and S.~Kim,
%``Thermal interquark potentials for bottomonium using NRQCD from the HAL QCD method,''
[arXiv:2112.09092 [hep-lat]].

%%%%%%%%%%%%%%%%%%%%%%%%%%%%%%%%%%%%%%%%%%%%%%%%%%%%%%%%%%%
%% Wilson loops
%%%%%%%%%%%%%%%%%%%%%%%%%%%%%%%%%%%%%%%%%%%%%%%%%%%%%%%%%%%

\bibitem{Hasenfratz:2001hp}
A.~Hasenfratz and F.~Knechtli,
%``Flavor symmetry and the static potential with hypercubic blocking,''
Phys. Rev. D \textbf{64}, 034504 (2001)
% doi:10.1103/PhysRevD.64.034504
[arXiv:hep-lat/0103029 [hep-lat]].

\bibitem{Bazavov:2019qoo}
A.~Bazavov \textit{et al.} [TUMQCD],
%``Determination of the QCD coupling from the static energy and the free energy,''
Phys. Rev. D \textbf{100}, no.11, 114511 (2019)
% doi:10.1103/PhysRevD.100.114511
[arXiv:1907.11747 [hep-lat]].

%%%%%%%%%%%%%%%%%%%%%%%%%%%%%%%%%%%%%%%%%%%%%%%%%%%%%%%%%%%

\bibitem{Hoying:2021mba}
D.~Hoying, A.~Bazavov, D.~Bala, G.~Parkar, O.~Kaczmarek, R.~Larsen, S.~Mukherjee, P.~Petreczky, A.~Rothkopf and J.~H.~Weber,
%``Static Potential At Non-zero Temperatures From Fine Lattices,''
[arXiv:2110.00565 [hep-lat]].

\bibitem{Steinbeisser:2018sde}
S.~Steinbei\ss{}er \textit{et al.} [TUMQCD],
%``Color screening in $2+1$ flavor QCD at large distances,''
PoS \textbf{Confinement2018}, 266 (2018)
% doi:10.22323/1.336.0266
[arXiv:1811.12846 [hep-lat]].

\bibitem{Petreczky:2021mef}
P.~Petreczky, S.~Steinbei\ss{}er and J.~H.~Weber,
%``Chromo-electric screening length in 2+1 flavor QCD,''
[arXiv:2112.00788 [hep-lat]].

%%%%%%%%%%%%%%%%%%%%%%%%%%%%%%%%%%%%%%%%%%%%%%%%%%%%%%%%%%%

\bibitem{Parkar:2021wpv}
G.~Parkar, D.~Bala, O.~Kaczmarek, R.~Larsen, S.~Mukherjee, P.~Petreczky, A.~Rothkopf and J.~H.~Weber,
%``In-medium static quark potential from spectral functions on realistic HISQ ensembles,''
[arXiv:2111.15437 [hep-lat]].
%0 citations counted in INSPIRE as of 01 Dec 2021

\bibitem{Burnier:2013nla}
Y.~Burnier and A.~Rothkopf,
%``Bayesian Approach to Spectral Function Reconstruction for Euclidean Quantum Field Theories,''
Phys. Rev. Lett. \textbf{111}, 182003 (2013)
% doi:10.1103/PhysRevLett.111.182003
[arXiv:1307.6106 [hep-lat]].

%%%%%%%%%%%%%%%%%%%%%%%%%%%%%%%%%%%%%%%%%%%%%%%%%%%%%%%%%%%

\bibitem{Petreczky:2018xuh}
P.~Petreczky, A.~Rothkopf and J.~Weber,
%``Realistic in-medium heavy-quark potential from high statistics lattice QCD simulations,''
Nucl. Phys. A \textbf{982}, 735-738 (2019)
% doi:10.1016/j.nuclphysa.2018.10.012
[arXiv:1810.02230 [hep-lat]].

\bibitem{Burnier:2014ssa}
Y.~Burnier, O.~Kaczmarek and A.~Rothkopf,
%``Static quark-antiquark potential in the quark-gluon plasma from lattice QCD,''
Phys. Rev. Lett. \textbf{114}, no.8, 082001 (2015)
% doi:10.1103/PhysRevLett.114.082001
[arXiv:1410.2546 [hep-lat]].

%%%%%%%%%%%%%%%%%%%%%%%%%%%%%%%%%%%%%%%%%%%%%%%%%%%%%%%%%%%

\bibitem{Bazavov:2014kva}
A.~Bazavov, Y.~Burnier and P.~Petreczky,
%``Lattice calculation of the heavy quark potential at non-zero temperature,''
Nucl. Phys. A \textbf{932}, 117-121 (2014)
% doi:10.1016/j.nuclphysa.2014.09.078
[arXiv:1404.4267 [hep-lat]].

\bibitem{Bala:2019cqu}
D.~Bala and S.~Datta,
%``Nonperturbative potential for the study of quarkonia in QGP,''
Phys. Rev. D \textbf{101}, no.3, 034507 (2020)
% doi:10.1103/PhysRevD.101.034507
[arXiv:1909.10548 [hep-lat]].

\bibitem{Bala:2021jcw}
D.~Bala, O.~Kaczmarek, R.~Larsen, S.~Mukherjee, G.~Parkar, P.~Petreczky, A.~Rothkopf and J.~H.~Weber,
%``The complex potential from 2+1 flavor QCD using HTL inspired approach,''
[arXiv:2112.00664 [hep-lat]].

%%%%%%%%%%%%%%%%%%%%%%%%%%%%%%%%%%%%%%%%%%%%%%%%%%%%%%%%%%%

\bibitem{Borsanyi:2015yka}
S.~Bors\'anyi, Z.~Fodor, S.~D.~Katz, A.~P\'asztor, K.~K.~Szab\'o and C.~T\"or\"ok,
%``Static $ \overline{\mathrm{Q}}\mathrm{Q} $ pair free energy and screening masses from correlators of Polyakov loops: continuum extrapolated lattice results at the QCD physical point,''
JHEP \textbf{04}, 138 (2015)
% doi:10.1007/JHEP04(2015)138
[arXiv:1501.02173 [hep-lat]].

\bibitem{Hart:2000ha}
A.~Hart, M.~Laine and O.~Philipsen,
%``Static correlation lengths in QCD at high temperatures and finite densities,''
Nucl. Phys. B \textbf{586}, 443-474 (2000)
% doi:10.1016/S0550-3213(00)00418-1
[arXiv:hep-ph/0004060 [hep-ph]].


\end{thebibliography}
\end{document}